\documentclass[journal]{IEEEtran}
\ifCLASSINFOpdf

\else
\fi

\usepackage{amsmath,amsfonts,amssymb,graphicx}
\usepackage[amsmath,thmmarks]{ntheorem}
\usepackage[tight,footnotesize]{subfigure}

\usepackage{url}
\usepackage{colortbl}
\usepackage{booktabs}
\usepackage{tabularx}
\usepackage{algorithm}
\usepackage{algorithmicx}
\usepackage{algpseudocode}
\newtheorem{mylemma}{Lemma}
\newtheorem{myDef}{Definition}
\newtheorem{mythm}{Theorem}

\usepackage{breakurl}
\usepackage{float}
\usepackage{multicol,multienum}
\usepackage{breqn}
\usepackage{stfloats}
\usepackage{multirow}
\usepackage{diagbox}
\usepackage{hyperref}
\usepackage{bm}
 \usepackage{cite}

\begin{document}

\title{ Dominant Dataset Selection Algorithms for Electricity Consumption Time-Series Data Analysis Based on Affine Transformation\thanks{Yi Wu, Yi Liu, and Jialiang Peng are with the School of Data Science
and Technology, Heilongjiang University, Harbin 150080, China. Syed Hassan
Ahmed is with the Department of Computer Science, Georgia Southern
University, Statesboro, GA 30460, USA. Ahmed A. Abd El-Latif is with
the Mathematics and Computer Science Department, Faculty of Science,
Menoufia University, Shebin El-Koom 32511, Egypt, and , and is with School of Information Technology and Computer Science, Nile University. Jialiang Peng is the corresponding author (e-mail: jialiangpeng@hlju.edu.cn, a.rahiem@gmail.com, AAbdelLatif@nu.edu.eg).}}

\author{Yi Wu, Yi Liu, Syed Hassan Ahmed, Jialiang Peng, Ahmed A. Abd El-Latif}

\markboth{Journal of \LaTeX\ Class Files,~Vol.~XX, No.~X, 201X}%
{Shell \MakeLowercase{\textit{et al.}}: Bare Demo of IEEEtran.cls for IEEE Journals}

\maketitle

\begin{abstract}

The explosive growth of time-series data, the scale of time-series data (TSD) suggests that the scale and capability of many Internet of Things (IoT)-based applications has already been exceeded. Moreover, redundancy persists in TSD due to correlation between information acquired via different sources. In this paper, we propose a cohort of dominant dataset selection algorithms for electricity consumption time-series data with focus on discriminating the dominant dataset that is small dataset but capable of representing the kernel information carried by time-series data with an arbitrarily small error rate less than $\varepsilon$. Furthermore, we prove that the selection problem of the minimum dominant dataset is an NP-complete problem. The affine transformation model is introduced to define as the linear correlation relationship between time-series data objects. Our proposed framework consists of the scanning selection algorithm with $O({n^3})$ time complexity and the greedy selection algorithm with $O({n^4})$ time complexity, which are respectively proposed to select the dominant dataset based on the linear correlation distance between timeseries data objects. The proposed algorithms are evaluated on the real electricity consumption data Harbin city in China. The experimental results show that the proposed algorithms not only reduce the size of extracted kernel dataset but also ensure the time-series data integrity in term of accuracy and efficiency.


\end{abstract}

\IEEEpeerreviewmaketitle
\section{Introduction}

\IEEEPARstart {A}{ccompanying} the growing popularity of smart grids and intelligent electric power networks is the generation and availability of large amounts of time-series data (TSD) in the power sector \cite{8281479}.
For example, the electricity consumption TSD in public institutions or private homes is continuously monitored via intelligent electric power systems using by Internet of Things (IoT) infrastructure. Information about electricity consumption is collected as  TSD from
smart sensors  and transmitted in real-time and analysed via IoT, so customers can be provided with the meaningful power usage data to help them utilize power more efficiently. In this sense, it is expected that the data analysis methods can be effectively used for advanced electricity planning and forecasting at different levels. In most cases, the assessment of such massive and dynamic TSD is time-consuming and resource intensive. More so, since TSD streams are continuous and decisions are often needed in real time \cite{8515030}. This has made efficient data extraction an important issue in IoT. Conventional extraction methods assume infinite computing and storage resources \cite{wu2014network}, which fail because electricity consumption TSD is associated with large-scale, low-value density, and strong correlation characteristics. Therefore, the more efficient data extraction methods are expected to process the massive TSD. For example, the approximate information extraction method using the summary data structure \cite{ref6} and the dimensional decomposition as well as recovery methods \cite{ref8}\cite{ref9} mainly focus on reducing the time complexity of algorithms. Consequently, advanced techniques for efficient data extraction are necessary.


Since the volume of TSD is always beyond the computation and storage capabilities of IoTs, one feasible solution is to dramatically reduce the amount of TSD involved in the computation. For example, several sampling based algorithms \cite{cheng2009sampling,ref16,ref15,he2015approximate} were proposed to sample a small portion of sensory data to answer queries based on the user-specified precision requirements. However, the characteristics and correlations of sensory data are neglected during the sampling procedure, and is impossible to accurately recover the original information. The data compression techniques were further proposed, such as linear regression based compression \cite{Tseng}, \cite{Deligiannakis}, source coding based compression \cite{5462034}, information entropy based compression \cite{4927470} have also been proposed. Moreover, the temporal demands imposed by the decompression process further complicates these approaches.

Recently, the usability theory was introduced in \cite{ref11} to analyze the big data issues. Similarly, the ${\left( {\varepsilon ,{\rm{ }}\delta } \right)}$ approximation theory \cite{ref2} was further proposed to select the high quality data related to big data, including the sample selection, the mathematical solver ${\left( {\varepsilon ,{\rm{ }}\delta } \right)}$ for a given problem, and the dynamic sample maintenance. According to the ``Do More with Less'' strategy for big data when the big data processing exceeds the computation and storage capacities, thence small data needs to be processed from big data.
Based on such a strategy, the several methods \cite{ref3,ref17-7,ref18-8,cheng2017extracting}
have been proposed to select the dominant datasets from big sensory data in wireless sensor networks.
The selected dominant datasets are applied as the small-scale datasets on which the data query operations can be completed under the given precision constraints. However, most of the above-mentioned methods lack efficient data correlation analysis needed for different big data application scenarios. More importantly, the real-time performances of the existing dominant dataset selection methods are often neglected or poorly accounted for. The above facts motivate our investigation of new dominant data selection methods to efficiently deal with massive TSD. Earlier efforts in  \cite{cheng2013varepsilon} and \cite{li2014approximate}  show there are both temporal and spatial correlation relationships between the massive TSD because the physical world always varies continuously in space
and time. To a certain extent, such strong correlations  in TSD accompany high data redundancy, i.e., majority of information carried by large-scale TSD can be represented
by a small-scale dataset referred to as a dominant dataset of TSD. Obviously, the information processing on a dominant dataset instead of the original massive TSD can significantly reduce the costs involved in computation, storage, and transmission.

In this work, we investigate how to select a dominant dataset in order to efficiently represent TSD. More specifically, in this paper,  we select a dominant dataset from the electric power consumption time-series data to  support the  analysis and management of TSD effectively.
Table \ref{tab-1} shows the real electric power consumption data in each time window. Table \ref{tab-1} can also be regarded as a TSD model that is an $m$ by $n$ matrix $X$ where $m$ and $n$ denote the sampling time points and the number of users in the given time window, respectively. Such a data matrix $X$ has high redundancy due to the strong correlations between the consumption data of consumers with similar living habits.
Thus, another data matrix $Y$ with much smaller dimensionality is expected to represent the matrix $X$.
In other words, for any given error rate $\varepsilon$, a small matrix $Y$ with the size of $m*k$ ($k \ll n$) can be selected as a dominant dataset if the information carried by the matrix $Y$  is compared to that by the original matrix $X$ with the information error rate being not more than $\varepsilon$.
In order to ensure the real-time performance of dominant dataset selection, we establish {a linear} reduction function $f: {x_i} \to {x_j}, (i,j = 1,2, \cdots ,n)$ if there is a linear correlation relationship between the column vectors $x_i$ and $x_j$ in matrix $X$.
Therefore,  $x_j$  can be represented by  ${f(x_i)\doteq x_j'}$
when the information difference  between  $x_j$ and $x_j'$ can satisfy the  requirement of error rate  being not more than $\varepsilon$.
Furthermore, it is assumed that ${f(x_2) \doteq x_1'}$, ${f(x_2) \doteq x_4'}$ and ${f(x_3) \doteq x_5'}$, where each error rate between $x_1$ and $x_1'$,  $x_4$ and $x_4'$,  $x_5$ and $x_5'$  is not more than $\varepsilon$. It means that the information carried by the ``large dataset'' ${ \left\{ {{x_1},{\rm{ }}{x_2},{\rm{ }}{x_3},{\rm{ }}{x_4},{\rm{ }}{x_5},{\rm{ }}{x_6}} \right\}}$  can be represented by a ``small dataset'' ${\left\{ {{x_2},{\rm{ }}{x_3},{\rm{ }}{x_6}} \right\}}$.
In essence, it also reflects the idea that the information processing  technology on the ``large dataset'' can be dealt on the ``small dataset''. In order to reduce the computational cost as much as possible, the optimal goal of seeking a  dominant dataset is to minimize its size. Meanwhile, the costs of computation and storage can be reduced by processing a  dominant dataset instead of the original massive TSD.
In the experiments, the  dominant dataset is selected to represent the kernel power consumption information of all users under the constraint of  error rate $\varepsilon$.
The main contributions of this paper are described as follows.

\begin{itemize}
\item[$\bullet$] We define the concept of dominant dataset for TSD, formalize the dominant dataset selection problem for TSD, and prove that the minimum dominant dataset is an NP-complete problem.

\item[$\bullet$] Based on the affine relation theory, an affine transformation model is applied as the reduction function to solve the linear correlation computational problem in TSD. In addition, using the proposed reduction, functions that can be dynamically updated by the increasing TSD to maintain the information processing adequately are presented.

\item[$\bullet$] Measuring the linear correlation between TSD is a key problem in selecting dominant datasets. Definitions of the affine linear correlation and the least square linear correlation are also presented. Further, the rigours of selecting appropriate dominant dataset of TSD that meet the requirements of information error rate based on the proposed correlation measures are also presented.

\item[$\bullet$] We propose a scanning selection algorithm (SSA) and the greedy selection algorithm (GSA) to determine dominant datasets based on the constraint of  ${\left( {\varepsilon ,{\rm{ }}\delta } \right)}$-solver. Finally, extensive experimental analyses are employed to validate the performance of the proposed algorithms in terms of both information representation accuracy and dominant dataset size.

\end{itemize}

\begin{table}[!htbp]
\begin{center}
\caption{Examples of electric power consumption data (kwh per-user)}
\begin{tabularx}{0.95\linewidth}{cXXXXXXX}%
\toprule
\diagbox{Times}{Users} 
&$x_1$ & $x_2$ & $x_3$ & $x_4$ & $x_5$&$x_6$ &$\cdots$ \\
\midrule
1:00&1.60&9.60&5.53&20.20&16.59 &13.90&$\cdots$\\
\hline
2:00&1.88&11.28&4.90&23.56&16.20 &16.20&$\cdots$\\
\hline
3:00&2.32&13.60&6.81&18.69&14.90&10.30&$\cdots$\\
\hline
4:00&4.32&11.50&9.10&17.69&13.90&14.60&$\cdots$\\
\hline
$\vdots$&$\vdots$&$\vdots$&$\vdots$&$\vdots$&$\vdots$&$\vdots$&$\vdots$\\

\bottomrule
\end{tabularx}
\label{tab-1}
\end{center}
\end{table}

The rest of this paper is organized as follows: Section \ref{sec-2} describes the related works. Section \ref{sec-3} introduces the dominant dataset definition, including proof that the selection problem of the minimum dominant dataset is an NP-complete problem. Section \ref{sec-4} presents the reduction function of the dominant dataset selection based on affine linear transformation. Section \ref{sec-5} provides the linear correlation measure definitions. Section \ref{sec-6} elaborates the proposed dominant dataset selection algorithms. The experimental results are presented and discussed in Section \ref{sec-7}. Finally, Section \ref{sec-8} concludes this paper.

\section{Related Works}
\label{sec-2}

Recent studies adopt data mining techniques to analyze electricity consumption and extract valuable information for the benefit of customers, utility companies, etc.
Data mining
techniques are mostly used to study and improve issues related to electricity
consumption patterns \cite{en11030683}.
In addition, several approaches related to
clustering  large-scale TSD have been recently proposed in \cite{rakthanmanon2013,ding2015yading,capo2017efficient}.
The information retrieved via data mining techniques are further used as input parameters to
forecast the electricity consumption based on regression, neural network, support vector machine, etc \cite{Kaytez2015Forecasting}.
However, to the best of our knowledge, none of the above methods apply dominant datasets to detect patterns of electricity  consumption from massive TSD collected by IoTs.
Therefore, this work intends to provide a reliable and accurate dominant dataset selection method as the basis for
these data mining algorithms dealing with massive TSD.

At present, most methods employed in high-dimensional variable selection require various assumptions to guarantee statistical properties required for low error rate and large power \cite{CIS-244161}. These methods have been applied for simultaneously selecting important variables and estimating their effects in high-dimensional statistical inference. Notwithstanding the complexity and difficulty of choosing a proper statistical model have also prevented these approaches from being widely used in practice even though most have the elegant theoretical properties \cite{Jing-Shiang} of these methods. For example, the stepwise regression is a method of fitting regression models in which the selection of predictive variables is carried out by an automatic procedure \cite{GVK022791892}. In each step, a variable is considered for addition to or subtraction from the set of explanatory variables based on some pre-specified criteria. Although the stepwise regression \cite{Chen08,tm08516,Ing} is a simple and powerful model selection method, it may not be a good choice in instances with large number of predictors and a relatively small number of observations, i.e. because the stepwise regression to select models with many spurious predictors. It indicates that the stepwise regression methods may be sensitive to specific model assumptions derived from linear regression models despite consistent properties that are theoretically justified. More importantly, the existing stepwise regression methods fail to maintain the dynamic update of the new time-series data arrival to select the dominant data in real time. The above facts motivate our effort to develop the dominant dataset selection algorithms for time-series data without statistical models.

Some dominant dataset selection methods \cite{ref6,ref8,ref9,ref7,ref20,ref10} were proposed in several areas, such as tradition database, data stream, wireless sensor network, etc. However, none of them are adapted for the high-quality information extraction associated with massive TSD \cite{ref19}. In order to reduce computational cost, a typical method using the summary data structure was proposed in \cite{ref6} to select a small-scale dataset as the approximate sampling dataset from a large-scale dataset. However, it is difficult for the selected sampling dataset to control the error of information extraction. The wavelet function was applied in \cite{ref7} to decompose the stream data to obtain approximate data query results. Based on both the coupling characteristics of stream data and the multi-level wavelet decomposition, the multi-stream compression methods were further proposed in \cite{ref8,ref9}. These methods offer 2 to 4 times better compression ratio than the traditional wavelet compression method \cite{ref7}. However, they fail to work well for the massive TSD analysis due to the high computational cost associated with Harr wavelet decomposition and recovery. {Meanwhile, in spite of its use of Discrete Fourier transform was used to analyse the coupling relation between the stream data, the method in \cite{ref10} failed to account for the effect of historical data for information extraction. Improvements in the highlighted methods mainly focus on reducing the time complexity without considering the scale of TSD are still not obvious for the computational efficiency of massive TSD.

Currently, representative methods \cite{ref11},\cite{ref1,ref12,ref13} have been proposed to support  the big data compression  approaches without the decompression computation. The original datasets are compressed in advance and then related computational operations can be done directly on the compressed datasets. These approaches reduce the size of original dataset to lower the computational cost, but they fail to solve the  online computational problems that accompany real-time TSD. The  ${\left( {\varepsilon ,{\rm{ }}\delta } \right)}$ approximate computing principle of big data was proposed in \cite{ref16},\cite{ref15},\cite{ref20,ref1,ref14} where the computational problem in the small dataset is solved by randomly extracting a small-scale dataset from the big dataset in accordance with the ${\left( {\varepsilon ,{\rm{ }}\delta } \right)}$ principle. However, the sampling probability for individual data is very low based on the random sampling procedure. Therefore, the ${\left( {\varepsilon ,{\rm{ }}\delta } \right)}$  principle is  suitable for the macro-data analysis rather than the micro-data query.
Although the dynamic statistics of partial TSD was proposed in \cite{ref4} based on the affine transformation theory, it failed to represent the complete  TSD information well. Unlike the existing methods, in this study, we implement the real-time data processing on the dominant dataset using the small-scale dominant dataset from large-scale TSD based on the given accuracy of information representation.

\section{Problem Definition}
\label{sec-3}

Time-varying objects in a timeline form the data sequence called time-series data (TSD), which we refer to as the time-series sample objects. Herein, a matrix ${X_{m \times n}}$  represents as $m\times n$ TSD objects where there are  $n$   sample objects  composed of  $m$ observation times per-object.

\subsection{Dominant Dataset}
A dominant dataset of the $n$ sample objects $\{{x_1},{\rm{ }}{x_2},...,{\rm{ }}{x_{\rm{n}}}\}$ is defined as follows.

\begin{myDef}\textbf{\emph{(Dominant Dataset).}}\label{defi-1}
{Assuming that ${X = \{ {\rm{ }}{x_i}|{x_i} \in \mathbb{R},1 \le i \le n\}}$ is  a  finite time-series dataset,
and let   $P = \{ {\rm{ }}{p_i}|{p_i} \in \mathbb{R},1 \le i \le k\}$ be another dataset,
${P \subset X}$, $|P| \ll |X|~(k \ll n)$. If there is a reduction function $f$  during the time period $T$ that can meet the requirement of  $f(P) \doteq Y_P$ and $X = Y_P \cup {{P}}$, then $P$ is defined as a dominant dataset of  $X$  based on the function $f$.	} \end{myDef}

\begin{myDef}\textbf{\emph{(${\bm{\left( {\varepsilon ,{\rm{ }}\delta } \right)}}$-solver)}}.\label{defi-2}
{Given the parameters $\varepsilon~(\varepsilon\geqslant 0)$ and $\delta~(0\leqslant \delta\leqslant 1) $,  the function $f$  is established by the correlation between the sample objects in $X$. A small dataset $P$ can be selected as a dominant
dataset of $X$ by the function $f$ such that the information representation problem of  $X$ can be solved in the small dataset $P$ instead of $X$ under the  condition that the probability of information extraction error being larger than  $\varepsilon$ is less than $\delta $. This solution condition of dominant dataset selection problem is defined as ${\left( {\varepsilon ,{\rm{ }}\delta } \right)}$-solver. If $\delta =0$, it means that the information extraction error  is less than $\varepsilon$. In such case, the solution condition is defined as $\varepsilon $-solver.}
\end{myDef}

\begin{myDef}\textbf{\emph{(Correlation Distance).}}\label{defi-3}
In order to measure the degree of correlation between $X$ and $Y$, the correlation distance between $X$  and $Y$ is defined as ${\mathfrak D}{\left( {{{X}},{{Y}}} \right)}$.
Based on ${\left( {\varepsilon ,{\rm{ }}\delta } \right)}$-solver, an element of dominant dataset can be determined if the condition  ${\mathfrak D}{(X,Y)}  \le  \varepsilon $  is met where the correlation distance is commutative, i.e. ${\mathfrak D}{(X,Y)}={\mathfrak D}{(Y,X)}$.
\end{myDef}

\begin{myDef}\textbf{\emph{(Central Object and Target Object).}}\label{defi-4}
Assuming that $X$ and $P$ are sample datasets (${P \subset X}$), there is a reduction function $f$ that can establish both $f(P) \doteq Y_P$ and $X = Y_P \cup {{P}}$  to be established. if the function $f$ meets  the requirement of ${{\left( {\varepsilon ,{\rm{ }}\delta } \right)}}$-solver, $P$ is a dominant dataset of $X$ based on
${\left( {\varepsilon ,{\rm{ }}\delta }\right)}$-solver during  the time period $T$.
{Each sample object in $P$  is defined as  a central object, and each sample object in  $Y_P$ is defined as a target object.}
\end{myDef}

\subsection{Dominant Dataset Selection}

In this work, the information representation  problem of large-scale TSD is transformed to a small-scale data processing problem that needs to meet the accuracy requirement of the information extraction  in the original  time-series dataset.

\begin{myDef}\textbf{\emph{(${\bm{\left( {\varepsilon ,{\rm{ }}\delta } \right)}}$-Dominant Dataset)}}\label{defi-5}
Given a sample dataset ${X = \{ {x_1},{x_2}, \ldots ,{x_n}\}}$ and the constraint  ${\left( {\varepsilon ,{\rm{ }}\delta } \right)}$-solver in time period $T$,
a dominant dataset ${P = \{ {p_1},{p_2}, \ldots, {p_k}\}}$ exists subject to the constraint  ${\left( {\varepsilon ,{\rm{ }}\delta } \right)}$-solver. It requests the existence of both the  reduction function $f$ and the correlation distance ${\mathfrak D_f}{\left( {{{X}},{{P}}} \right)}$ meeting the requirement of ${\left( {\varepsilon ,{\rm{ }}\delta } \right)}$-solver to  establish  $f(P) \doteq Y_P$ and $X = Y_P \cup {{P}}$ (${P \subset X}$, ${|P| \ll |X|}$). Here, $P$ is defined as the ${{\left( {\varepsilon ,{\rm{ }}\delta } \right)}}$-dominant dataset.
\end{myDef}

\begin{mylemma}\label{mylemma1}
If there are multiple dominant datasets $\mathbb{P}=\{{P_1},{P_2}, \ldots ,{P_w}\},~ w \in {\mathbb{ N^+}}$ in  a time-series dataset $X$ corresponding to the different  reduction functions $F{\rm{ }} = \left\{ {f_1,f_2,{\rm{ }}...f_w} \right\}$ with the determined  ${\left( {\varepsilon ,{\rm{ }}\delta } \right)}$-solver, then a minimum dominant dataset (a dominant dataset with the minimum size) $P_{min}$  exists in $\mathbb{P}$.\label{lemma1}\end{mylemma}

\textbf{Proof.} When the different  reduction functions  are determined by ${\left( {\varepsilon ,{\rm{ }}\delta } \right)}$-solver, it can be seen that the dominant dataset, namely, the  dominant dataset selection result of $X$, is not unique based on Definition \ref{defi-5}. Therefore, there exists multiple dominant datasets $\mathbb{P}=\{{P_1},{P_2}, \ldots ,{P_w}\},~w \ge 1$ for $X$ corresponding to the different  reduction functions $F{\rm{ }} = \left\{ {f_1,f_2,{\rm{ }}...f_w} \right\}$. Therefore, it is bound to exist a dominant dataset $P_{min}$ with the minimum size in $\mathbb{P}$. $\blacksquare$

\begin{myDef}\textbf{\emph{(The Inclusion Problem of Dominant Dataset).}}\label{defi-6}
Let $U_p = \left\langle {{{X,\mathbb{P}}}} \right\rangle $,
where $X = \left\{ {x_1,{\rm{ x_2}},...,{\rm{ }}x_d} \right\}$ is  a time-series dataset and  the several subsets of $X$ make up the set
$\mathbb{P}=\{ {P}|{P} \subset X\},~\left| \mathbb{P} \right| = {2^{\left| X \right|}}$.
The inclusion problem of  minimum dominant dataset is defined to select an  element $P_{min}$ in $\mathbb{P}$ to make $P_{min}$  be the minimum dominant dataset of $X$ based on a given reduction function $f$.
\end{myDef}

\begin{myDef}\textbf{\emph{(The Selection Problem of Minimum Dominant Dataset).}}\label{defi-7}
Let $C_q=\left\langle {X,F} \right\rangle$, 
where $X = \left\{ {x_1,{{ }x}_2,...{\rm{ }}x_d} \right\}$ is a time-series dataset and $F{\rm{ }} = \left\{ {f_1,f_2,{\rm{ }}...f_w} \right\}$  is the available reduction function set for $X$ based on  ${\left( {\varepsilon ,{\rm{ }}\delta } \right)}$-solver.
The consequent dominant datasets are $\mathbb{P}'=\left\{ {P_1,{\rm{ }}P_2,...{\rm{ }}P_w} \right\}$, where ${f_i}({P_i}) \buildrel\textstyle.\over= Y_{P_i}~(\forall {{{P}}_i} \in {{\mathbb{P}'}},~1 \le i \le w)$ , then the problem of selection of minimum dominant dataset is reduced to selection ${{{P}}_i} \in {\mathbb{P}}'$  under the condition that the size of $P_{i}$ is minimum.
\end{myDef}

\begin{mylemma}The inclusion problem of the 
 dominant datasets is NP-complete problem.\label{lemma2}\end{mylemma}
\textbf{Proof.} Based on the  graph theory,  a dominating set \cite{Gross2013} in an undirected graph $G$ with a set of $T$ of vertices such that every vertex in $G$ is either in $T$ or connected to a vertex of $T$ by an edge, or both. The dominating set problem is to input an undirected graph $G$ and a number $k$, and determine whether there is a dominating set with $k$ vertices.
The dominating set problem is a classical NP-complete problem in computational complexity theory \cite{Garey1979}.
According to Definition \ref{defi-1}, a dominant dataset $P$ is a subset of the time-series dataset $X$ that can be approximately represented by $P$ based on the reduction function $f$. Based on  Definition \ref{defi-6}, $X$ can be abstractly considered as an undirected graph $G$.  ${x_i} \in X$ and $f_i\in F$ can further be represented as a vertex and an edge in  $G$, respectively. Consequently, the problem of finding a  dominant dataset $P$ of $X$ in this work is equivalent to the dominating set problem in the  graph theory. Therefore, the inclusion problem of  dominant datasets is NP-complete problem. $\blacksquare$

\begin{mythm}{The solution of the inclusion problem and that of the  selection problem for the minimum dominant dataset are  equivalent.}\label{th1}\end{mythm}
\textbf{{Proof.}}	
For the  reduction function set $F$, it needs to prove that a solution for  the minimum dominant dataset  selection problem of $C_q$ corresponds to a solution for the  minimum dominant dataset inclusion problem  of $U_p$ needs to be proven. That is, given $S(C_q)$ as one solution instance of $C_q$, one solution instance, $S (U_p)$, for $U_p$, can be obtained, and vice versa. $\blacksquare$   

For $S({C_q}) =  > S({U_p})$, according to the requirement of dominant dataset, given  $S\left( {{C_q}} \right)$, there is
a reduction function $f$ to produce  $S\left( {{C_q}} \right)$  according to the requirement of dominant dataset.
Each element 
in $S\left( {{C_q}} \right)$ belongs to the dataset $X$ based on the definition of ${C_q}$. Moreover, $S\left( {{C_q}} \right)$ is a subset of $\mathbb{P}$ and since the reduction function $f$ can produce $S\left( {{C_q}} \right)$ as the minimum dominant dataset of $\mathbb{P}'$ then $S\left( {{C_q}} \right)$  is also a solution of the dominant dataset inclusion problem for ${U_p}$.

For ${S\left( {U_p} \right) =  > S\left( {C_q} \right)}$, according to the definition of ${U_p}$, given $S(U_p)$, $S(U_p)$ is a defined subset of the dataset $\mathbb{P}$  according to the definition of $U_p$.
A reduction function $f$ can produce the minimum dominant dataset $P_{min}$ based on the definition of $C_q$, and all elements of $P_{min}$  belong to  $\mathbb{P}$, namely, ${{P_{min}} \subseteq \mathbb{P}}$. Therefore, $S(U_p)$ is a solution of the minimum dominant dataset selection problem for $C_q$ based on the reduction function $f$.

In summary, it is proven that  the solution of the inclusion problem and that of the selection problem for the minimum dominant dataset are equivalent.

\begin{mythm}The selection problem of the minimum domination dataset is NP-complete problem.\label{th2}\end{mythm}
\textbf{{Proof.}}
According to Lemma \ref{lemma2} and Theorem \ref{th1}, the inclusion problem of the dominant datasets is NP-complete, and both the inclusion problem and the selection problem of the minimum dominant dataset are
equivalent. Therefore, the selection problem of the minimum domination dataset is an NP-complete problem. $\blacksquare$

\section{Mathematical Foundation}
\label{sec-4}

Inspired by the work of Saket, \textit{et. al} in \cite{ref4}, we introduce the affine relation theory to implement the information extraction of TSD based on the choice of a suitable affine transformation function. Affine transformation is a nonsingular linear transformation between two vector spaces. We use the affine relation model as the reduction function to construct the target object vector space of TSD for the dominant dataset selection.

\subsection{Affine Relation Model}

In this work, the affine relation model  is defined as ${S = P \times A + B}$, where
$S$ and $P$ represent a sample object matrix and central object matrix of the TSD  respectively (see Definition \ref{defi-4}),
 $A$ is  a coefficient matrix and $B$ denotes a residual matrix.

It is assumed that  a 2-dimensional affine relation can be expressed as ${S_{m \times 2}} = {P_{m \times 2}} \times {A_{2 \times 2}} + {B_{m \times 2}}$, where both the sample object matrix ${S_{m \times 2}}= ({y_1},{y_2})$ and the central object matrix ${P_{m \times 2}}=({x_1},{x_2})$ are $m$-row by 2-column matrices, $A = ( {a_1},{a_2})$ is a 2-row by 2-column coefficient matrix,  and $B = ({b_1},{b_2})$ is an $m$-row by 2-column constant matrix. Without loss of generality, a 2-dimensional affine relation model can be defined as follows:
\begin{equation}
\label{eq-1}
\resizebox{.95\hsize}{!}{$
	\left( {\begin{array}{*{20}{c}}
  {{y_1}\left( 1 \right)}&{{y_2}\left( 1 \right)} \\
   \vdots & \vdots  \\
  {{y_1}\left( {\text{m}} \right)}&{{y_2}\left( {\text{m}} \right)}
\end{array}} \right) = \left( {\begin{array}{*{20}{c}}
  {{x_1}\left( 1 \right)}&{{x_2}\left( 1 \right)} \\
   \vdots & \vdots  \\
  {{x_1}\left( {\text{m}} \right)}&{{x_2}\left( {\text{m}} \right)}
\end{array}} \right)\left( {\begin{array}{*{20}{c}}
  {{a_{11}}}&{{a_{12}}} \\
  {{a_{21}}}&{{a_{22}}}
\end{array}} \right) \\+ \left( {\begin{array}{*{20}{c}}
  {{b_{1}(1)}}&{{b_{1}(2)}} \\
   \vdots & \vdots  \\
  {{b_{m}(1)}}&{{b_{m}(2)}}
\end{array}} \right)$}
\end{equation}
Eq. \eqref{eq-1} can be further considered as the  affine  transformation from $x_1$, $x_2$  to $y_1$, $y_2$,  as shown in Eq. \eqref{eq-2}.
\begin{equation}
\label{eq-2}
\resizebox{.91\hsize}{!}{$
 {\begin{cases}
{{y_1}\left( i \right) = {a_{11}} \times {x_1}\left( i \right) + {a_{21}} \times {x_2}\left( i \right) + {b_{i}(1)}}\\
{{y_2}\left( i \right) = {a_{12}} \times {x_1}\left( i \right) + {a_{22}} \times {x_2}\left( i \right) + {b_{i}(2)}}
\end{cases}}({i = 1,2, \ldots ,m})$}
\end{equation}

Next, we extend both  $S$ and  $P$ to $m$-row
by $n$-column matrices. The $n$-dimensional  affine relation model can be represented as  ${S_{m \times n}} = {P_{m \times n}} \times  {A_{n \times n}} + {B_{m \times n}}$, where  $A_{n \times n}$ is the transformation  coefficient matrix and $B_{m \times n}$ is the residual matrix.
Let ${S_{m \times n}}= ({y_1},{y_2}, \cdots ,{y_m})$, ${P_{m \times n}} = ({x_1},{x_2}, \cdots ,{x_m})$, ${{\rm{A}}_{n \times n}}  = \left( {{a_1},{a_2}, \ldots ,{a_n}} \right)= {\left( {{a_{ij}}} \right)_{n \times n}}~(i,j = 1,2, \ldots ,n)$,  ${B_{m \times n}} = ({b_1},{b_2}, \cdots ,{b_m})$, and then
\begin{equation}
\label{eq-3}
{{y_i} = \left( {\mathop \sum \limits_{j = 1}^n {a_{ji}}{x_i}_{}} \right) + {b_i}}, i = 1,2, \ldots ,m.
\end{equation}

\subsection{Affine Transformation Function}

The affine relation between  the sample object matrix $S$ and the central object matrix $P$ is assumed as ${\Re:(A,B)}$, and let the corresponding matrix $R = \left( {\begin{array}{{c}}
  {{A_{n \times n}}} \\
  {{B_{1 \times n}}}
\end{array}} \right)$, where there are $(n + 1) \times n$  elements in the matrix $R$ .
Based on the affine relation model ${S = P \times A + B}$, let ${{P'}} = ( P,1_m)$ and then $S = P' \times R$. Thus, Eq. \eqref{eq-3} can be further expressed as presented in Eq. \eqref{eq-4}.
\begin{equation}
\label{eq-4}
	\resizebox{.95\hsize}{!}{$
\left( {\begin{array}{*{20}{c}}
  {{y_1}\left( 1 \right)}&{{y_2}\left( 1 \right)}& \cdots &{{y_n}\left( 1 \right)} \\
  {{y_1}\left( 2 \right)}&{{y_2}\left( 2 \right)}& \cdots &{{y_n}\left( 2 \right)} \\
   \vdots & \vdots & \cdots & \vdots  \\
  {{y_1}\left( m \right)}&{{y_2}\left( m \right)}& \cdots &{{y_n}\left( m \right)}
\end{array}} \right) = \left( {\begin{array}{*{20}{c}}
  {{x_1}\left( 1 \right)}&{{x_2}\left( 1 \right)}& \cdots &{{x_n}\left( 1 \right)}&1 \\
  {{x_1}\left( 2 \right)}&{{x_2}\left( 2 \right)}& \cdots &{{x_n}\left( 2 \right)}&1 \\
   \vdots & \vdots & \cdots & \vdots & \vdots  \\
  {{x_1}\left( m \right)}&{{x_2}\left( m \right)}& \cdots &{{x_n}\left( m \right)}&1
\end{array}\;} \right)\left( {\begin{array}{*{20}{c}}
  {{a_{11}}}&{{a_{12}}}& \cdots &{{a_{1n}}} \\
  {{a_{21}}}&{{a_{22}}}& \cdots &{{a_{2n}}} \\
   \vdots & \vdots & \cdots & \vdots  \\
  {{a_{n1}}}&{{a_{n2}}}& \cdots &{{a_{nn}}} \\
  {{b_1}}&{{b_2}}& \cdots &{{b_n}}
\end{array}} \right)$}
\end{equation}

Since that the central object matrix $P$ aims to represent  different target objects in TSD,
it can also ensure the nonlinear correlation relation between the sample object vectors when the time window of TSD $m$ is very large. That is to say, $P$ can have full column rank within a large time window.
Due to $S = P' \times R$, there is a pseudo-inverse matrix of $P'$ to establish $R = pinv(P') \times S$ where $pinv(P')$ denotes the  pseudo-inverse matrix of $P'$. Meanwhile, the equation
${pinv(P') = {\left( {{P^{'T}} \times P'} \right)^{ - 1}} \times {P^{'T}}}$ holds.
When the dimensionality of the sample object vector is $m$,  the equation ${R_m} = {\left( {P_m^{'T} \times P_m^'} \right)^{ - 1}} \times {P^'}_m^T \times {S_m}$  also holds.
When $m$ increases to $m+1$ in Eq. \eqref{eq-4} (namely, an additional observation of TSD with increasing the time window), it increases  a row  in the matrices $P'$ and $S$ respectively, as shown in Eq. \eqref{eq-5}.
\begin{equation}\label{eq-5}
\begin{gathered}
  P{'_{m + 1}} = \left( {\begin{array}{*{20}{c}}
  {P{'_m}} \\
  {p{'_{m + 1}}}
\end{array}} \right),~
  {S_{m + 1}} = \left( {\begin{array}{*{20}{c}}
  {{S_m}} \\
  {{s_{m + 1}}}
\end{array}} \right) \\
\end{gathered}
\end{equation}
where $p'_{m + 1} = \left( {{x_1}\left( {m + 1} \right),{x_2}\left( {m + 1} \right), \cdots ,{x_n}\left( {m + 1} \right),1} \right)$ and ${s_{m + 1}} = \left( {{y_1}\left( {m + 1} \right),{y_2}\left( {m + 1} \right), \cdots ,{y_n}\left( {m + 1} \right)} \right)$.
Therefore, ${R_{m + 1}} = {\left( {P_{m + 1}^{'T} \times P_{m + 1}^'} \right)^{ - 1}} \times {P^'}_{m + 1}^T \times {S_{m + 1}}$. {It can be derived as}
\begin{equation}\label{eq-6}
\begin{aligned}
{R_{m + 1}}&= {\left( {{I_n} + {{\left( P_{m}^{'T}P'{_m} \right)}^{ - 1}}p_{m + 1}^{'T}p'{_{m + 1}}} \right)^{ - 1}} \\
 &\times \left( {{R_m} + {{\left( {P'_m}^TP{'_m} \right)}^{ - 1}}{p^{'T}_{m + 1}}{s_{m + 1}}} \right).
\end{aligned}
\end{equation}
Please see the Appendix for the proof of Eq. \eqref{eq-6}.

For the time-series data, the sample object matrix $S$ can be represented by the central object matrix $P$ based on the affine relation model. 
It means that the affine relation can be used to extract target object information based on the central  object matrix $P$. $P'$ are actually collected in real time as shown in Eq. \eqref{eq-5}, and the transform matrix $R$ can be also real-time computed  by Eq. \eqref{eq-6}. Therefore, the sample objects can be efficiently represented by $P'$, $R$ matrices. Herein, $R$  also means the transformation function in this work.

As shown in Eq. \eqref{eq-6}, $R$  can be dynamically updated by the continuous arrival of TSD samples.
The  dynamic update of $R$ guarantees the maintenance of the continuous information extraction from large-scale time-series datasets.

\section{Information Loss and Linear Correlation Distance}
\label{sec-5}

In accordance with  the accuracy requirement of information representation, we study  how to select the dominant dataset from TSD  under the error constraint. Given the error constraint $\varepsilon$, the dominant dataset is selected  by the affine transformation function, and then the dominant dataset is  evaluated by ${{\rm{(}}\varepsilon ,\delta )}$-solver to analyze the usability of information representation.

\subsection{Information Loss}

Based on the definition of affine relation model in  Eq. \eqref{eq-4} and the transformation function $R$ in Eq. \eqref{eq-6}, the dataset $S=(y_1,y_2,\cdots,y_n)$  can be reconstructed as $S'=(y'_1,y'_2,\cdots,y'_n)$ based on the  dominant object dataset $P$ and $S' = P \times R$. Therefore, the
 information loss  between  $S$ and $S'$ is defined as $E = |S - S'|$.
\begin{equation}\label{eq-7}
	\resizebox{.95\hsize}{!}{$
		\begin{aligned}
	E = \left( {\begin{array}{*{20}{c}}
  {|{y_1}\left( 1 \right) - y{'_1}\left( 1 \right)|}&{|{y_2}\left( 1 \right) - y{'_2}\left( 1 \right)|}& \cdots &{|{y_n}\left( 1 \right) - y{'_n}\left( 1 \right)|} \\
  {|{y_1}\left( 2 \right) - y{'_1}\left( 2 \right)|}&{|{y_2}\left( 2 \right) - y{'_2}\left( 2 \right)|}& \cdots &{|{y_n}\left( 2 \right) - y{'_n}\left( 2 \right)|} \\
   \cdots & \cdots & \cdots & \cdots  \\
  {|{y_1}\left( m \right) - y{'_1}\left( m \right)|}&{|{y_2}\left( m \right) - y{'_2}\left( m \right)|}& \cdots &{|{y_n}\left( m \right) - y{'_n}\left( m \right)|}
\end{array}} \right)
\end{aligned}$}
\end{equation}
According to Definition \ref{defi-2},  $E$ is subject to the condition that the information loss of the target objects is less than $\varepsilon$ or the error extraction proportion of the target objects exceeding  $\varepsilon$ is not more than $\delta$.
Therefore, the information loss  of target
objects is also determined by the $(\varepsilon, \delta)$-solver constraint.

Given  $S=(y_1,y_2,\cdots,y_n)$  and $S'=(y'_1,y'_2,\cdots,y'_n)$ of $m$ observation times,
the root mean square error (RMSE) vector between $S$ and $S'$ is defined as Eq. \eqref{eq-8},
\begin{equation}\label{eq-8}
	E_{\_RMSE} = \left( {E_{\_RMSE_1}},E_{\_RMSE_2}, \cdots ,E_{\_RMSE_n} \right)
\end{equation}where
$E_{\_RMSE_j} = \sqrt {\frac{{\mathop \sum \nolimits_{i = 1}^m {{\left( {{y_j}\left( i \right) - y{'_j}\left( i \right)} \right)}^2}}}{m}} \;\;\left( {1 \leqslant j \leqslant n} \right)$.	
Thus, the information loss between  $S$ and $S'$ can be analyzed based on $E_{\_RMSE}$.

\subsection{Linear Correlation Distance}

\begin{myDef}\textbf{\emph{(Linear Correlation Distance, LCD).}}\label{defi-8} Assuming that there are  pair matrices of ${{\cal P}}=(x_1,x_2)$ and ${{\cal S}}=(y_1,y_2)$, ${\mathfrak D_{{{LCD}}}}{( {{{\cal P}},{{\cal S}}} )}$ is defined as the linear correlation distance between $\cal P$ and $\cal S$.\end{myDef}

Based on the degree of ${\mathfrak D_{{{LCD}}}}{( {{{\cal P}},{{\cal S}}} )}$ as it tends to 0,
LCD can be used to measure the linear correlation relationship between $\cal P$ and $\cal S$. ${\mathfrak D_{{{LCD}}}}{( {{{\cal P}},{{\cal S}}} )}$ is further applied to determine whether $\cal S$ can be affine transformed by $\cal P$ to implement the information representation. If ${\mathfrak D_{{{LCD}}}}{( {{{\cal P}},{{\cal S}}} )} \le \varepsilon $, it means that LCD meets the requirements for information extraction accuracy. When the linear correlation distance ${\mathfrak D_{{{LCD}}}}{ {{({\cal P}},{\rm{\cal S}}} )} \le \varepsilon$ meets the requirement of ${{\rm{(}}\varepsilon ,\delta )}$-solver, it implies that the couple vectors $y_1$, $y_2$ can be linearly represented by  the couple vectors $x_1$, $x_2$ based on the affine transformation function.

Inspired by the work in \cite{ref4},  we further introduce the terms: public object vector, central object vector, and sample object vector in this work, where it is assumed that
\begin{itemize}
\item[$\bullet$]$u$ represents a public  object vector,

\item[$\bullet$]$p$ represents a central  object vector,

\item[$\bullet$]$v$ represents a target  object vector,

\item[$\bullet$]$F_{aff}$ represents an affine transformation function,

\item[$\bullet$]$R $ represents an affine transformation matrix.
\end{itemize}
In \cite{ref4}, two  $m$-by-2 matrices are respectively defined as the pivot pair matrix $(u,p)$  and the sequence pair matrix $(u,v)$. The generating procedure for the pivot pair matrix is conducted by Affine Clustering Algorithm\cite{ref4}. Furthermore, the covariance for all the pivot pair matrices are computed to determine the affine transformations between each sequence pair matrix and one of the pivot pair matrices. Here, the pivot pair matrix and the sequence pair matrix play the role of the affine transformation as shown in Fig. \ref{fig-1}. Although the above procedure can be used for computing $(u,p)$   and $(u,v)$,  in our proposed methods, we simplify the generating procedure for $(u,p)$   by selecting a random object vector $u$   as a public object vector for a central vector  $p$ to find the target object $v$  based on the proposed affine transformation function between   ${{\cal P} = ( {u,p} )}$  and $ {\cal S} = (u,v)$   in this work.
As the same definition in \cite{ref4}, ${{\cal P} = ( {u,p} )}$  and $ {\cal S} = (u,v)$  are represented as the pivot vector pair and the target vector pair respectively, in order to determine the linear correlation distance between ${{\cal P} }$  and $ {\cal S} $ based on the condition of  ${\mathfrak D_{{{LCD}}}}{( {{{\cal P}},{{\cal S}}} } )\le \varepsilon $.

Therefore, let the sample object vector set ${X = \{ {x_1},{x_2},...,{x_n}\} }$, and ${{{\cal P}} = (u,p),{{\cal S}} = (u,v)}$, $\exists u \in X  $,  $\forall p \in X$, $\forall v \in X$.
If the linear correlation distance between $\cal P$ and $\cal S$  satisfies the condition of ${\mathfrak D_{{{LCD}}}}{( {{{\cal P}},{{\cal S}}} } )\le \varepsilon $, the linear correlation distance between $p$  and $v$  is less than $\varepsilon $. Let  a vector $ p \in \cal P$ be a central vector for a target vector $ v \in \cal S$ based on an affine transformation function ${F_{aff}:p \to v}$, 
and then all  central vectors form the dominant dataset ${P = \{ {p_1},{p_2}, \ldots ,{p_k}\}}$.
As shown in Fig. \ref{fig-1}, the public object vector $u$  is selected to form both the pivot vector pair $\cal P$ and the target vector pair $\cal S$, therefore, the vector $v$ in $\cal S$  can be computed by the vector $p$ in  $\cal P$ based on the transformation function $F_{aff}$.
\begin{figure}[htpb]
	\centering
	\includegraphics[width=0.95\linewidth]{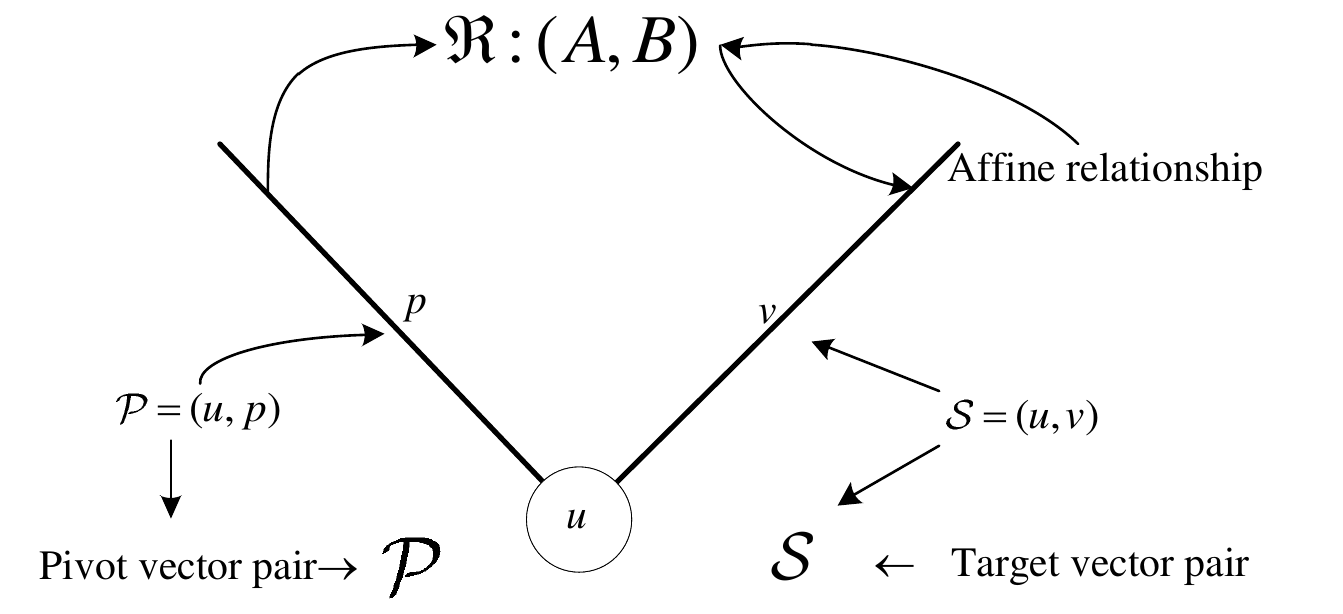}
	\caption{{Procedure for generating the pivot pairs \cite{ref4}}.}
	\label{fig-1}
\end{figure}

Assuming that the central object dataset ${\cal P}$ and the sample object dataset ${\cal S}$, the two linear distance measures are introduced as follows:

\begin{itemize}
\item[$\bullet$]\textbf{ Affine linear correlation distance (AFF)}:
Let the affine relation $ \Re_{AFF}:(A,B)$ be used for the pair matrices ${\cal P}$ and $ {\cal S}$.
The dataset ${\cal S'}_{AFF}$ can be obtained by ${\cal S'}_{AFF} = {{A}} \times {\cal P} + {{B}}$ to approximatively represent ${\cal S}$. The linear correlation distance $\mathfrak D_{LCD}({\cal P},{\cal S})$ can be represented by  the affine linear correlation ${\mathfrak D_{{AFF}}}({\cal P},{\cal S})$ as shown in Eq. \eqref{eq-9}

\begin{equation}\label{eq-9}
{\mathfrak D_{AFF}{({\cal P},{\cal S})}\buildrel \Delta \over= ||{{\cal S}_{}}\!-\!{{\cal S'}_{AFF}}||}.
\end{equation}
When $\mathfrak D_{AFF}{({\cal P},{\cal S})} \le \varepsilon $, it can meet  the requirement  of the $(\varepsilon ,\delta )$-solver constraint. { It is obvious that the vectors $u,v$ in ${\cal S}$  can be linearly represented by the vectors $u,p$ in ${\cal P}$.}

\item[$\bullet$]\textbf{{Least-squares linear transformation distance (LS)}}: As similar as the affine relation distance, let the least-squares transformation \cite{introduction} $\Re_{LS}:(A,B)$ be used for the pair matrices ${\cal P}$ and $ {\cal S}$ . The dataset ${{\cal S'}_{LS}}$ can be also obtained by ${\cal S'}_{LS} = {{A}} \times {\cal P} + {{B}}$ to approximatively represent ${\cal S}$.  The linear correlation distance $\mathfrak D_{LCD}({\cal P},{\cal S})$ can be represented by the least-squares linear distance ${\mathfrak D_{_{LS}}}({\cal P},{\cal S})$ as shown in Eq. \eqref{eq-10}

\begin{equation}\label{eq-10}
	{{\mathfrak D_{_{LS}}}{\left( {{\cal P},{\cal S}} \right)} \buildrel \Delta \over = \left| {\left| {{\cal S} - {\cal S'}_{LS}} \right|} \right|}.
\end{equation}
When ${\mathfrak D_{LS}{({\cal P},{\cal S})} \le \varepsilon}$,  it can also meet the requirement  of the $(\varepsilon ,\delta )$-solver constraint to linearly represent the vectors $u,p$ in { ${\cal S}$ by the vectors $u,p$ in ${\cal P}$.}

\end{itemize}

\section{Dominant Dataset Selection Algorithms}
\label{sec-6}

Theorem \ref{th2} proves that the dominant dataset selection problem is an NP-complete problem.
According to Definition \ref{defi-8}, the dominant dataset is selected under the condition that the linear correlation distance is subject to the constraint of ${(\varepsilon ,\delta )}$-solver. Based on  Theorem \ref{mylemma1}, we construct $k$ linear correlation groups from a given TSD  matrix $X_{m\times n}$.
The  distance between a central target  object $S_p$ in each group and any other  sample object $S_v$ in the same group is required to meet the requirement of the ${(\varepsilon ,\delta )}$-solver constraint. In fact, it hopes to find
the  dominant dataset $P$ with the  minimum size $k$ can meet the requirement of  the ${(\varepsilon ,\delta )}$-solver constraint.

\subsection{Scanning Selection Algorithm}

We propose the scanning selection algorithm (SSA) to select the dominant dataset based on the linear correlation distance  measure with the constraint of ${(\varepsilon ,\delta )}$-solver. The basic idea of  SSA  is described as follows:

\begin{itemize}
\item[$\bullet$] A linear independent object pair { ${\cal P}=(u,p)$} is selected from the sample object dataset $X$ based on  the sequential object  order of  $X$.   The  target  object $v$ is identified by traversing $X$ to form the target object pair ${\cal S}=(u,v)$ that is subject to the constraint ${\mathfrak D_{LCD}(\cal P,\cal S)} \le \varepsilon$.
\item[$\bullet$]  
   The central object $p$ is added into the dominant dataset $P$, and  the affine transformation matrix $\Re$ is formed by its corresponding affine function  ${F_{aff}:p \to v}$.

\item[$\bullet$]  The above procedure is repeated until the identification of the central objects and the target objects for all  objects in $X$ is completed to give the final dominant dataset $P$.

\end{itemize}

The proposed SSA is executed using  Algorithm \ref{alg1}.
The related variables are initialized in Line 1, and the corresponding target object is identified by  the  sequential central object in Lines 2 through 18.
This iterative process  that  the central object is selected from unidentified target objects is described in Lines 5 through 15, where the central object is identified based on the constraint of $(\varepsilon ,\delta )$-solver.
The identified central object $p$ and the corresponding transformation function coefficient matrix $(A_j, b_j)$ are added to the dominant dataset $P$ and the transformation coefficient matrix set $A$ respectively, in Line 16. In Line 17, the next unidentified target object is orderly selected  as the central object  until $X$ becomes an empty set. Finally, the  dominant set $P$ and the affine
transformation coefficient matrix set $A$ are output from SSA.

\begin{algorithm}
\caption{Scanning Selection Algorithm (SSA)}\label{alg1}
{\bf Input:}
TSD matrix ${X_{m \times n} = \{ x_1, x_2,..., x{_{n}}\} }$ and ${(\varepsilon ,\delta )}$-solver.\\
{\bf Output:}
The dominant dataset $P$ and
the  affine transformation coefficient matrix set ${A}$.
		\begin{algorithmic}[1]	
	\State $P \leftarrow \emptyset$, $A \leftarrow \emptyset$, $i \leftarrow 1$, $X \leftarrow {X_{m \times n}}$, ${n_\delta } \leftarrow 0$;%
	\While{${X \ne \emptyset }$}
	\State {$p \leftarrow {x_i}$, $\exists {u} \in X$,  {${\cal P} \leftarrow (u,p)$}, $X \leftarrow X - \left\{ {{x_i}} \right\}$;}
	\State ${P_{v_i}} \leftarrow \emptyset$, ${A_{p_i}} \leftarrow \emptyset $;
	\For{$j \in \left[ {1,|X|} \right], j \ne i$}
	\State $\exists {x_j} \in X,{{v}} \leftarrow {x_j}$, ${\cal S} \leftarrow (u,v)$;
	\If{${\mathfrak D_{{{LCD}}}}{\left( {{\cal P},{\cal S}} \right)} \le \varepsilon $}
	\State {${{P_{v_i}} \leftarrow {P_{v_i}} \cup \{ v\}, {A_{p_i}} \leftarrow {A_{p_i}} \cup \{ ({A_j},b_j)\} }$;}
	\Else
	 \If{{${n_\delta } < [\left( {|P| + |A|} \right) \times \delta {\rm{]}}$}}
	\State ${{P_{v_i}} \leftarrow {P_{v_i}} \cup \{ v\}, {A_{p_i}} \leftarrow {A_{p_i}} \cup \{ ({A_j},b_j)\} }$;
	\State ${n_\delta } \leftarrow {n_\delta } + 1$;
\EndIf
	\EndIf
	\EndFor
	\State $P \leftarrow P \cup \{ p\} ,A \leftarrow A \cup  {A_{p_i}} $;
	\State $X \leftarrow X - \{P_{v_i}\},i \leftarrow i+1$;
    \EndWhile
	\State \Return $P$, $A$.
\end{algorithmic}
\end{algorithm}

In this work, we introduce the directed graph structure to describe the dominant selection relationship between the central objects and the target objects under conditions imposed via the ${(\varepsilon ,\delta )}$-solver. Let a dominant relationship be a directed graph $G = (X,E)$, where ${X = \{ x_1,x_2,...,x_{{n}}\} }$ is the vertex set of $G$ representing all sample objects, $E$ is the  directed edge set of $G$,  and a directed edge in $E$ representing an affine transformation relationship between a central object $x_i \in X$ and a target object $x_j \in X, j \ne i$
that meets the constraint of ${(\varepsilon ,\delta )}$-solver.
As shown in Fig. \ref{fig-2}, we further illustrate a sample for the dominant dataset selection process based on SSA.
In Fig. \ref{fig-2}(a), the dominant relationship is denoted as the directed graph structure $G=(X,E)$, where
$X = \{{x_1},{x_2},{x_3},{x_4},{x_5},{x_6}\}$ and
${E = \{{\left\langle {{x_1},{x_2}} \right\rangle ,\left\langle {{x_2},{x_1}} \right\rangle ,\left\langle {{x_2},{x_4}} \right\rangle ,\left\langle {{x_3},{x_5}} \right\rangle } }\}$.
Initially, $x_1$ is  sequentially selected as the central object $p$ from $X$ to construct  the target object pair $(u,p)$.
And then, $x_2$ is orderly selected as the object $v$ in order to construct the object pair $(u,v)$.
From Fig. \ref{fig-2}(a), we can see that the linear correlation distance  between   $x_1$ and $x_2$ meets the constraint of ${\mathfrak D_{{{LCD}}}}{( {x_1,x_2} )} \le \varepsilon $.
Therefore, $x_2$ is selected as the target object for the central object $x_1$, and then the transformation coefficient matrix set $A$   is updated with the affine function $F_{aff} $ until the central object $x_1$ does not satisfy the constraint.
In Fig. \ref{fig-2}(b),
$x_3$ is orderly selected as the central object from the remaining target objects $\{{x_3},{x_4},{x_5},{x_6}\}$, and then $x_5$ is selected as the target object dominated by $x_3$ under the constraint of ${\mathfrak D_{{{LCD}}}}{( {x_3,x_5} )} \le \varepsilon $. However, there are no target objects for $x_4$ when $x_4$ is orderly selected as the central object from the remaining target objects $\{{x_4},{x_6}\}$ as shown in Fig. \ref{fig-2}(c). Similarly, Fig. \ref{fig-2}(d) shows that $x_6$ is selected as the central object to complete the  dominant dataset selection process.
Finally, the dominant object dataset $P = \{{x_1},{x_3},{x_4},{x_6}\}$  and the affine transformation function coefficient matrix  set ${{A = }}\{ ({{{A}}_2},{b_2}),({{{A}}_5},{b_5})\} $ for $\{x_2,x_5\}$ are output as the results based on SSA.
\begin{figure}[htbp]
	\centering
	\includegraphics[width=0.90\linewidth]{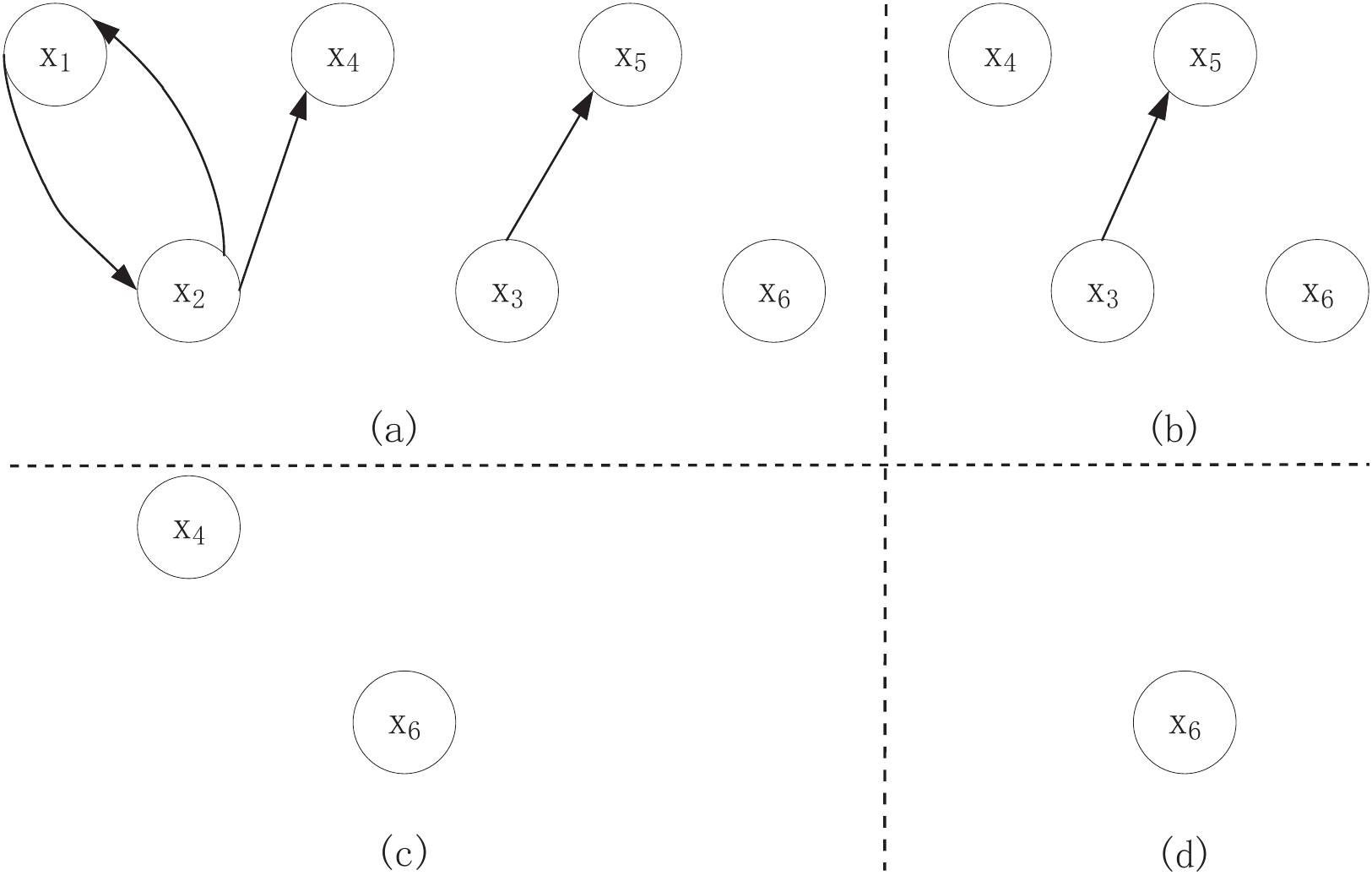}
	\caption{{A sample for the dominant dataset selection process  based on SSA.}}
	\label{fig-2}
\end{figure}

In the dataset $X$ containing $n$ target objects, the central objects are identified successively to constitute the dominant dataset  based on SSA.
The time complexity of Algorithm \ref{alg1} is the sum of running-time costs for each statement in it. In Line 1, the running-time cost is $O(m\times n)$  for initializing the related variables. In Lines 2 through 18, it is a while loop to allow the scanning selection of dominant dataset to be executed repeatedly based on the condition $ X \ne \emptyset $ is satisfied. It is assumed that the computational cost of the while loop is  $ O(n \times T)$ , where $T$  is the running-time cost of the statements in Lines 3 through 17. In Lines 3 and 4, it is constant time for the assignment of the related variables. In Lines 5 through 15, it is assumed that the running-time cost is $ O(n \times Z)$ for the for loop in which the dominant data objects are identified based on the constraint of $ \left( {\varepsilon ,{\rm{ }}\delta } \right)$-solver, where  $Z$ is the computational cost of executing statements in Lines 6 through 14. Except for the statement in Line 7, a constant time is required to execute the statements in Line 6 and Lines 10 through 13. To judge the inequality of ${\mathfrak D_{{{LCD}}}}{\left( {{\cal P},{\cal S}} \right)} \le \varepsilon $  in Line 7, we simplify to compare $\varepsilon ^2$  to the inner product of  ${{\cal P},{\cal S}}$ with the time complexity of  $O(n)$  in the implementation of Algorithm \ref{alg1}.
In Lines 16 and 17,, it is also a constant time to update $P$, $A$, etc. The computational cost $Z = O(1 + n) = O(n)$ and the running-time cost for the for loop is  $O(n \times Z) = O({n^2})$. Therefore,  $T = O(1 + {n^2} + 1) = O({n^2})$. So, the computational cost of the while loop is  $ O(n \times T) = O({n^3})$ and the into the time complexity of SSA is  $O(m\times n) + O(n^3)$, namely, $O(n^3)$.

\subsection{Greedy Selection Algorithm}
In the dominant dataset, it is expected that greedy selection of the dominant dataset   maximizes the coverage of the sample object dataset $X$ within the constraint of ${(\varepsilon ,\delta )}$-solver is expected, such that the size of $P$ is minimised.

Given a sample object dataset $X$, there is an object subset $P_v, P_v\subseteq X$ that contains the target objects corresponding to a central object $p \in X$ meets the requirement of ${(\varepsilon ,\delta )}$-solver.
The  greedy  strategy of selecting a central object is to choose $p$ from $X$ to obtain the corresponding target object subset $P_v$ meeting the following condition:
 \begin{equation}\label{eq-13}
{\mathop {argmax}\limits_{p} (|\sigma (P_v,X)|),{\text{ }}s.t.{\text{ }}{P_v} \subseteq X,~p \in X}
\end{equation}	
where  ${\sigma {{(P_v,X)}}}$ represents to form the subset $P_v$ that contains the target objects determined by the central object $p$  based on ${(\varepsilon ,\delta )}$-solver.

Therefore, the basic idea of the proposed dominant dataset greedy selection algorithm (GSA) is described as follows:
\begin{itemize}
\item[$\bullet$]   The  linear correlation distance $\mathfrak D_{LCD}$ between any two objects in $X$ are computed to find which paired  objects can meet the requirement of $\mathfrak D_{LCD} \le \varepsilon $. According to the above greedy selection strategy, the central object $p$ supporting the largest size of the target object subset  is added into the dominant set $P$, the corresponding target objects are added into $P_v$, and $X$ is updated as $X=X - \{ p\}  - {P_v} $. In addition, the affine transformation matrix $\Re$ is further formed by its corresponding affine function  ${F_{aff}:p \to P_v}$.
\item[$\bullet$]  The above procedure is repeated  until the dataset $X$ becomes empty. Finally, the dominant dataset greedy  selection of  $X$ is completed to give the final dominant dataset $P$.
\end{itemize}

Meanwhile, according to the definition of $(\varepsilon ,\delta )$-solver, the  proportion  of the target objects  corresponding a central object $p$ in $P_v$ dissatisfying with the condition $\mathfrak D_{LCD} \le \varepsilon $ can be less than $\delta $. The vector $E\__{RMSE}$ (see Eq. \eqref{eq-8}) is used to select a central object when  the above-mentioned situation arises.
Therefore,
the  target object $v$   in the dataset $X$ can be also identified by the condition of the minimum $E_{RMSE}$ in the proposed GSA when $\mathfrak D_{LCD} > \varepsilon $.
Given a sample object dataset $X$ and  a  target object subset $P_{v'}$   corresponding to the central object $p$ do not meet the condition of $\mathfrak D_{LCD} \le \varepsilon $, the current greedy selection operation  for a target object $v'$ from  $X$ 
should meet the following condition:
\begin{equation}\label{eq-14}
\begin{aligned}
 &\mathop {argmin}\limits_{v'}(E\__{RMSE_{v'}}),\;\\
 &s.t.{\text{ }}\sigma '(P_{v'},X),{\text{ }}p \in X,\;{P_{v'}} \subseteq X,\;v' \in {P_{v'}}
\end{aligned}
\end{equation}
where ${\sigma {{'(P_{v'},X)}}}$ represents to the dataset $P_{v'}$ that contains the target object $v'$  corresponding to the central object $p$ that does not meet the  condition of $\mathfrak D_{LCD} \le \varepsilon $ in the dataset $X$.

Algorithm \ref{alg2} executes the proposed GSA outlined as follows.
The related variables are initialized in Line 1.  In Line 2-12,
the linear correlation distance between  any two target objects in $X$ is computed to determine whether ${\mathfrak D_{{{LCD}}}}{\left( {{{\cal P}},{{\cal S}}} \right)} \le \varepsilon $ can be held
to form the dominant relationship, and the affine coefficient matrices are merged into the set $Ap$.
Next, the dominant object is further selected and the affine transformation function is constructed in  Line 13-24.
Based on the dominant  relationship in $X$, the  greedy selection of dominant objects is conducted to choose the central object $p$  supporting the largest size of the target
object subset $P_v$ in Line 15. In Line 16, the central object $p$ and the corresponding transformation function  coefficient matrix $Ap_p=(A_p, b_p)$  are added to  the central object set $P$ and the coefficient matrix set $A$, respectively. $X$ is updated as $X=X - \{ p\}  - {P_v} $ in Line 17.
If the proportion of target objects dissatisfying the condition of $\mathfrak D_{LCD} \le \varepsilon $ is smaller than $\delta$ in Line 18,  the target object $p$ is also allowed to be selected
to support the object $v'$ from the unrecognized target objects in $X$  under the condition of the smallest $E\_RMSE_{v'}$ in Line 19. The corresponding transformation function coefficient matrix $({A_{v'}},b_{v'})$ are joined into the coefficient  matrix  set $A$ in Line 20.
$X$ and $n_\delta $ are updated in Line 21-22. The above procedure in Line 14-24 is repeated to do until $X$ becomes empty. Finally, the
dominant set $P$ and the affine transformation coefficient matrix
set $A$ are output as the results  based on GSA.

\begin{algorithm}
	\caption{Greedy Selection Algorithm (GSA)}\label{alg2}
	{\bf Input:}
	TSD matrix  ${X_{m \times n} = \{ x_1, x_2,..., x{_{n}}\} }$ and ${(\varepsilon ,\delta )}$-solver.\\
	{\bf Output:}
	The dominant dataset ${P}$  and 	
	the affine coefficient matrix set $A$.	
	\begin{algorithmic}[1]	
		\State ${X \leftarrow {X_{m \times n}}}$, ${A_{p} \leftarrow \emptyset }$;
		\For   {$i \in \left[ {1,\left| X \right|} \right]$}
		\State {${{p} \leftarrow {x_i}}$, $\exists {u} \in X$, ${\cal P} \leftarrow (u,p) $, ${X_v} \leftarrow {{X}} - \left\{ {{x_i}} \right\}$;}
		\State ${A_p}_i \leftarrow \emptyset $;
		\For {$j \in \left[ {1,\left| {{X_v}} \right|} \right]$, $j \ne i$}
		\State ${{v}} \leftarrow {x_j}$, ${\cal S} \leftarrow (u,v)$;
		\If { ${\mathfrak D_{{ {LCD}}}}{\left( {{\cal P},{\cal S}} \right)} \le \varepsilon $}
		\State ${A_p}_i \leftarrow {A_p}_i\cup \{ ({A_j},b_j)\} $
		\EndIf
		\EndFor
		\State $A_p \leftarrow A_p \cup  {A_p}_i $
		\EndFor
		\State $P \leftarrow \emptyset$, $A \leftarrow \emptyset ,{P_v} \leftarrow \emptyset$, ${n_\delta } \leftarrow 0$;
		\While{${X \ne \emptyset }$}
		\State {$\mathop {argmax}\limits_{p} (\sigma (|P_v,X)|),{\text{ }}s.t.{\text{ }}{P_v} \subseteq X,~p \in X$}
		\State $P \leftarrow P \cup \{ p\}$, $A \leftarrow A \cup {A_p}_p$;
		\State $X \leftarrow X - \{ p\}$, $X \leftarrow X - {P_v} $;
		\While {${n_\delta } < \{ (|P| + |{A_p}|) \times \delta \}$~and~$X \ne \emptyset $}
	    \State {\resizebox{.82\hsize}{!}{$\mathop {argmin}\limits_{v'} (E\_{_{RMS{E_{v'}}}}),\;s.t.{\text{ }}\sigma '(P'_v,X),{\text{ }}p \in X,\;{P_{v'}} \subseteq X,\;v' \in {P_{v'}}$}};
		\State $A \leftarrow A \cup \{ ({A_{v'}},b_{v'})\} $;		
        \State $X \leftarrow X - {P_{v'}} $;
		\State ${n_\delta } \leftarrow {n_\delta } + 1$;		
		\EndWhile
		\EndWhile
		\State \Return $P$,~$A$.	
	\end{algorithmic}
\end{algorithm}

We also illustrate  a sample for the   dominant dataset selection process based on GSA as shown in Fig. \ref{fig-3}.
Similarly, Fig. \ref{fig-3}(a) shows the directed graph structure  $G=(X,E)$  where
$X = \{{x_1},{x_2},{x_3},{x_4},{x_5},{x_6}\}$ and	
${E = \{{\left\langle {{x_1},{x_2}} \right\rangle ,\left\langle {{x_2},{x_1}} \right\rangle ,\left\langle {{x_2},{x_4}} \right\rangle ,\left\langle {{x_3},{x_5}} \right\rangle } }\}$.
As shown Fig. \ref{fig-3}(a), the  object $x_2$ has 2 dominant relationships with the target objects $x_1$, $x_4$ .
Compared to other objects in $X$, there are the maximum  dominant relationships for $x_2$ in $X$ if $x_2$ becomes a central object.
Therefore, $x_2$ is firstly selected as the central object to add into the dominant dataset $P$ and $x_1$, $x_4$ are joined into the set $P_v$ based on GSA.
The  transformation function $(A_1,b_1),(A_4,b_4)$ of  $x_2$ is also computed to form the affine function coefficient matrix $A=\{(A_1,b_1),(A_4,b_4)\}$.
Next, the dataset $X$ is updated by $X \leftarrow X-\{x_1,x_2,x_4\}$ as shown in Fig. \ref{fig-3}(b). The object $x_3$ is selected as the central object to merge into the dominant dataset $P$, and the corresponding target object $x_5$ and the affine coefficient function $(A_5,b_5)$ are added into the  set $P_v$ and $A$, respectively. In  Fig. \ref{fig-3}(c), $X$ is further updated to $X=\{x_6\}$, and $x_6$ is added into $P$ because  $x_6$ is the last one object in  the current $X$. So far,
GSA  has been completed in the original $X$ to obtain the  dominant dataset $P=\{x_2,x_3,x_6\}$ and the affine function coefficient matrix set $A=\{(A_1,b_1),(A_4,b_4),(A_5,b_5)\}$ for the target objects $x_1$, $x_4$, $x_5$.

\begin{figure}[htpb]
	\centering
	\includegraphics[width=0.95\linewidth]{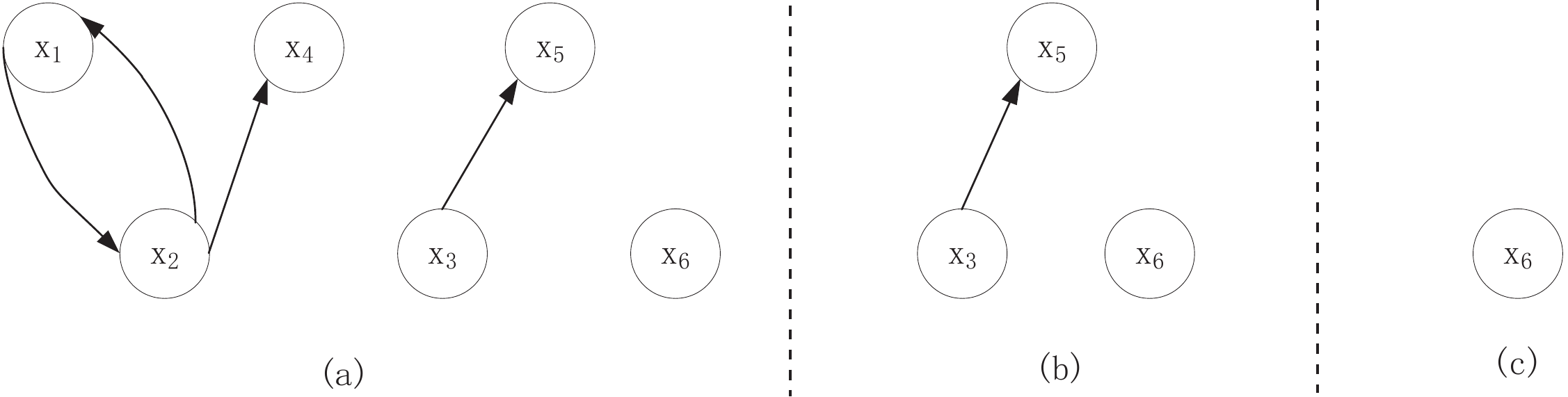}
	\caption{A sample for the dominant dataset selection process based on GSA.}
	\label{fig-3}
\end{figure}

The time complexity of Algorithm \ref{alg2} is the sum of running-time costs for each
statement in it. In Line 1, the running-time cost is $O(m\times n)$ for
initializing the related variables. In Line 2-12, it is a nested for loop to
determine the linear correlation distance between any two target objects in
$X$ based on the condition of ${\mathfrak D}_{LCD} \left( {{\cal P},{\cal S}}
\right) \le \varepsilon $. According to the time complexity analysis of Algorithm \ref{alg1}, ${\mathfrak D}_{LCD} \left( {{\cal P},{\cal S}}
\right)$ can be simplified to calculate within
the time complexity of $O(n)$. Therefore, the running-time cost of
statements in Line 2-12 of the nested for loop is $O(n^3)$. It is constant
time for the assignment of the related variables in Line 13. In Line 14-24, it
is a nested while loop to conduct the greedy selection of dominant dataset.
It is assumed that $O(n\times T)$ is the running-time cost  for the outer
while loop, where $T$ is the computational cost of executing the
statements in Line 15-23. The time complexity is $O(n)$ to select the central
object $p$ supporting the largest size of the target object subset $P_v
$ in Line 15. It is also constant time to execute the statements of Line
16-17.
For the inner while loop in Line 18-23, it is assumed that  $O(n \times Z)$ is the running-time cost for it where $Z$  is the computational cost of executing the statements in Line 19-22. The running-time cost is $O({n^2})$  for the statement Line 19 based on Eq. \eqref{eq-14} as the computational cost of $E_{_{RMSE}}$  is  $O({n^2})$, and it is constant time for the update of $A$, $X$, and $n_\delta$  in Line 20-22. Therefore, $Z = O({n^2} + 1) = O({n^2})$  and the running-time cost for the inner while loop is  $O(n \times Z) = O({n^3})$. So,   $T = O(n + 1 + {n^3}) = O({n^3})$ and the running-time cost for the outer while loop is $O(n \times T) = O(n \times {n^3}) = O({n^4})$. In summary, the total time
complexity of GSA is $O(m\times n) + O(n^3) + O(n^4)$, namely,
$O(n^4)$.

\subsection{Reconstruction of Target Object Dataset}

The algorithm SSA or GSA  outputs a dominant object dataset ${P = \left\{ {{p_1},{p_2}, \ldots ,{p_k}} \right\}}$  and a affine transformation coefficient matrix set ${A = \{{A_{{p_1}}},{A_{{p_2}}}, \ldots ,{A_{{p_k}}}\} }$ for the dataset $X$, where ${A_{p_i}\left( {1 \le i \le k} \right)}$   represents the affine transformation  coefficient matrix  for the target objects corresponding to the central object ${p_i}$  in  $P$. The  reconstruction method of target objects based on the dominant dataset is proposed in Algorithm \ref{alg3}.

\begin{algorithm}
	\caption{Reconstruction of Target Object Dataset}\label{alg3}
	{\bf Input:}
A dominant  dataset
${P = \left\{ {{p_1},{p_2}, \ldots ,{p_k}} \right\}}$,
the affine  transformation function $F_{aff}$,  and the affine transformation coefficient matrix
 set ${A = \{ A_{p_1},A_{p_2},...,A_{p_k}\} }$ for $P$.\\
	{\bf Output:}
	The reconstructed target object dataset $X_p$. 
	\begin{algorithmic}[1]	
		\State ${X_p = \emptyset }$;
		\For {{$i \in [1,k]$}}
	    \State ${{{q_i}} \leftarrow {F_{aff}}({A_{{p_i}}},{p_{i}})}$;
	    \State ${{X_p} \leftarrow  {X_p} \cup \{{q_i}\}}$;
		\EndFor
		\State \Return $X_p$.	
	\end{algorithmic}
\end{algorithm}

According to Algorithm \ref{alg3}, the reconstruction for the target dataset $X_p$ can be obtained  by the dominant dataset $P$ through computing the affine transformation function $F_{aff}$  with the  matrix set $A$. The time complexity for computing a target object ${q_i}$ is {$O(m \times n)$} based on the applied linear transformation function $F_{aff}$. Therefore, the time complexity of Algorithm \ref{alg3} is $O(m \times n \times k)$. Because  $m$ is always a constant and $k \le n$, the time complexity of target objects reconstruction  is $O( n^2)$.

The space complexity for the TSD matrix ${X_{m \times n}}$ is ${O\left( {m \times n}\right)}$.
When the size of the dominant dataset is $k$, the size of the target object dataset is $n-k$.
It means that the size of TSD storage space can be greatly reduced by only storing the dominant dataset in case that ${k \ll n}$. 
In addition, the reconstruction of target objects can be implemented by the the linear affine transformation function without high computational cost. Based on  Eq. \eqref{eq-6}, the next  ${R_{m + 1}}$ can also be recursively derived from $R_m$.
Therefore, it implies that the proposed algorithms  are also suitable to solve the related big data issues.

\section{Experimental Results And Analysis}
\label{sec-7}

We analyze the effectiveness  and efficiency of the proposed dominant dataset selection  methods in this experiment. Based on the ${(\varepsilon ,\delta )}$-solver, the performances of SSA and GSA algorithms are evaluated by
the affine linear correlation and the least squares linear correlation measurements, respectively. { In addition, the {reconstruction accuracy} of target objects is also analyzed in the experiment.}

\subsection{Experimental Setup}

The proposed algorithms are implemented using Python programming language on the Anaconda Navigator platform \cite{Anaconda}.
The experimental computing environment is a Windows PC with Intel i7-6770 CPU and 4GB RAM.
The experimental dataset is derived from the real  electricity consumption dataset of Harbin city in China collected in July 2013.
This dataset consists of more than 6 million users and 7-day power consumption data   per user at least, including
130 million power consumption records in total. Each consumption record contains the attributes of time, user number, power station number, power supply bureau number, and electricity consumption. According to the TSD model applied in this work, the user number, station number, and power supply bureau number attributes are redefined  as the user ID (namely, the target object ID). Therefore, the electricity consumption attribute becomes the most valuable information of TSD in the experiment. Table \ref{tab-2} presents a list of the description  of datasets
DS0, DS1, DS2, DS3 and DS4 extracted from the different power supply stations where dataset sizes increases from DS1 to DS4 datasets.

\begin{table}[htbp]
	\caption{THE DESCRIPTIONS OF THE EXPERIMENTAL DATASETS}
\begin{tabularx}{0.95\linewidth}{>{\hsize=0.2\hsize}X>{\hsize=0.1\hsize}X>{\hsize=0.1\hsize}c>{\hsize=0.2\hsize}c>{\hsize=0.7\hsize}X}
		\toprule[1pt]
		Dataset & Users & Days & Data records
		 & Description   \\
&($n$)&($m$)&(thousand)&\\
		\midrule[.5pt]
		DS0&1032&21 & 21.672&21-day consumption data for each user\\
		DS1&3000&7&21.000& 7-day consumption data for each user\\
		DS2&6000&7&42.000&7-day consumption data for each user\\
		DS3&9000&7&63.000&7-day consumption data for each user\\
	    DS4&12000&7&84.000&7-day  consumption data for each user\\
		
		\bottomrule[1pt]
	\end{tabularx}
	\label{tab-2}
\end{table}

We mainly focus on the effectiveness and efficiency aspects of the proposed algorithms in the experiment. First,
whether the size of  dominant dataset can be controlled by the constraints. The  size of dominant dataset meeting the ${(\varepsilon ,\delta )}$-solver constraint is an important objective in the experiment.
Let DSN\_ratio be the ratio of the size of dominant dataset to the number of sample objects.
It means that the smaller DSN\_ratio, the lower size of dominant dataset  achieved.
Herein, the optimal dominant dataset is decided by the analysis of DSN\_ratio. Second,
the execution efficiency of the proposed algorithms is further verified and analyzed when
 the reconstruction accuracy of  target objects is within an error range by the given constraint.
In the following experiments, we present the experimental results to analyze the above two aspects
with the changing parameters of the ${(\varepsilon ,\delta )}$-solver constraint and the size of experimental dataset.

\subsection{Performance Analysis of the Proposed Algorithms}

Based on the proposed SSA and GSA algorithms as well as AFF and LS measures, we implement the dominant dataset selection methods with the different measures, namely, the SSA\_AFF, SSA\_LS, GSA\_AFF, and GSA\_LS methods.
When the values  of error $\varepsilon$ are respectively set as 1\%, 3\%, 5\%, 8\%, and 10\% .
The changes in  DSN\_ratio achieved on the DS0 dataset by the above four methods in the experiment are plotted in Fig. \ref{fig-4}. The DSN\_ratio results  are also listed in
Table \ref{tab3}.
Fig. \ref{fig-4a} indicates that the GSA methods achieve lower DSN\_ratio  than the SSA methods  with the given errors based on the same measure.
Meanwhile, the AFF measurement results in  lower  DSN\_ratio than the LS measurement based on the same algorithm. Compared to other three methods, therefore,
the  smallest dominant dataset is achieved by the method GSA\_AFF.

We  use the expectation of $E_{\_RMSE} $  to measure the reconstruction accuracy of target objects in the experiment.
Table \ref{tab4} lists the mean RMSE results based on the above four methods with different $\varepsilon$.
Fig. \ref{fig-4b} also depicts the above mean RMSE results.
The mean RMSE of the GSA\_AFF method is generally less than that of other three methods as shown in Fig. \ref{fig-4b}.
It indicates that the proposed GSA\_AFF method achieves the best performance in terms of the size of dominant dataset and  the reconstruction accuracy.

\begin{table}[htbp]
\begin{center}
	\caption{THE DSN\_RATIO  WITH DIFFERENT $\varepsilon $.}
\begin{tabular}
{lllll}
\toprule
 $\varepsilon $&
SSA{\_}AFF&
SSA{\_}LS&
GSA{\_}AFF&
GSA{\_}LS \\
\midrule
1{\%}&
98.60{\%}&
99.50\%&
98.00\%&
99.50\% \\

3{\%}&
87.98{\%}&
95.74\%&
83.20\%&
93.60\%\\

5{\%}&
72.67{\%}&
86.43\%&
65.30\%&
82.30\% \\

8{\%}&
51.45{\%}&
70.06\%&
43.30\%&
62.80\%\\

10{\%}&
40.40{\%}&
58.72\%&
35.00\%&
50.80\% \\
\bottomrule
\end{tabular}
\label{tab3}
\end{center}
\end{table}

\begin{table}[htbp]
\begin{center}
	\caption{THE MEAN RMSE WITH DIFFERENT $\varepsilon $.}
\begin{tabular}{lllll}
\toprule
 $\varepsilon $&
SSA{\_}AFF&
SSA{\_}LS&
GSA{\_}AFF&
GSA{\_}LS \\
\midrule
1{\%}&
0.50&
0.48&
0.49&
0.42 \\
3{\%}&
1.50&
1.53&
1.50&
1.52 \\
5{\%}&
2.44&
2.60&
2.46&
2.62 \\
8{\%}&
3.96&
4.16&
4.00&
4.13 \\
10{\%}&
5.03&
5.17&
4.96&
5.20 \\
\bottomrule
\end{tabular}
\label{tab4}
\end{center}
\end{table}

\begin{figure}[htbp]
	\centering
	\subfigure[]{\includegraphics[width=0.48\linewidth]{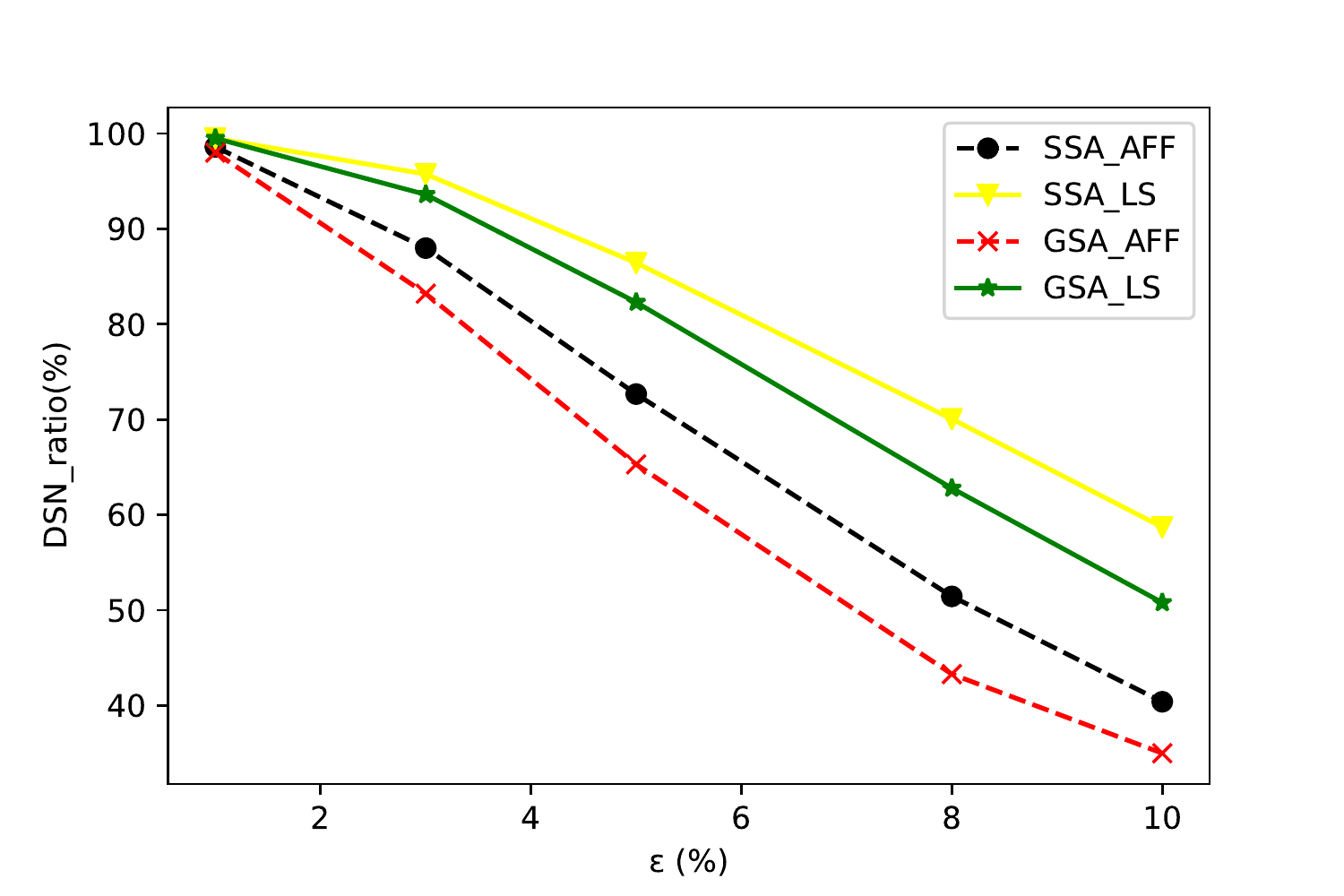}%
		\label{fig-4a}}
	\hfil
	\subfigure[]{\includegraphics[width=0.47\linewidth]{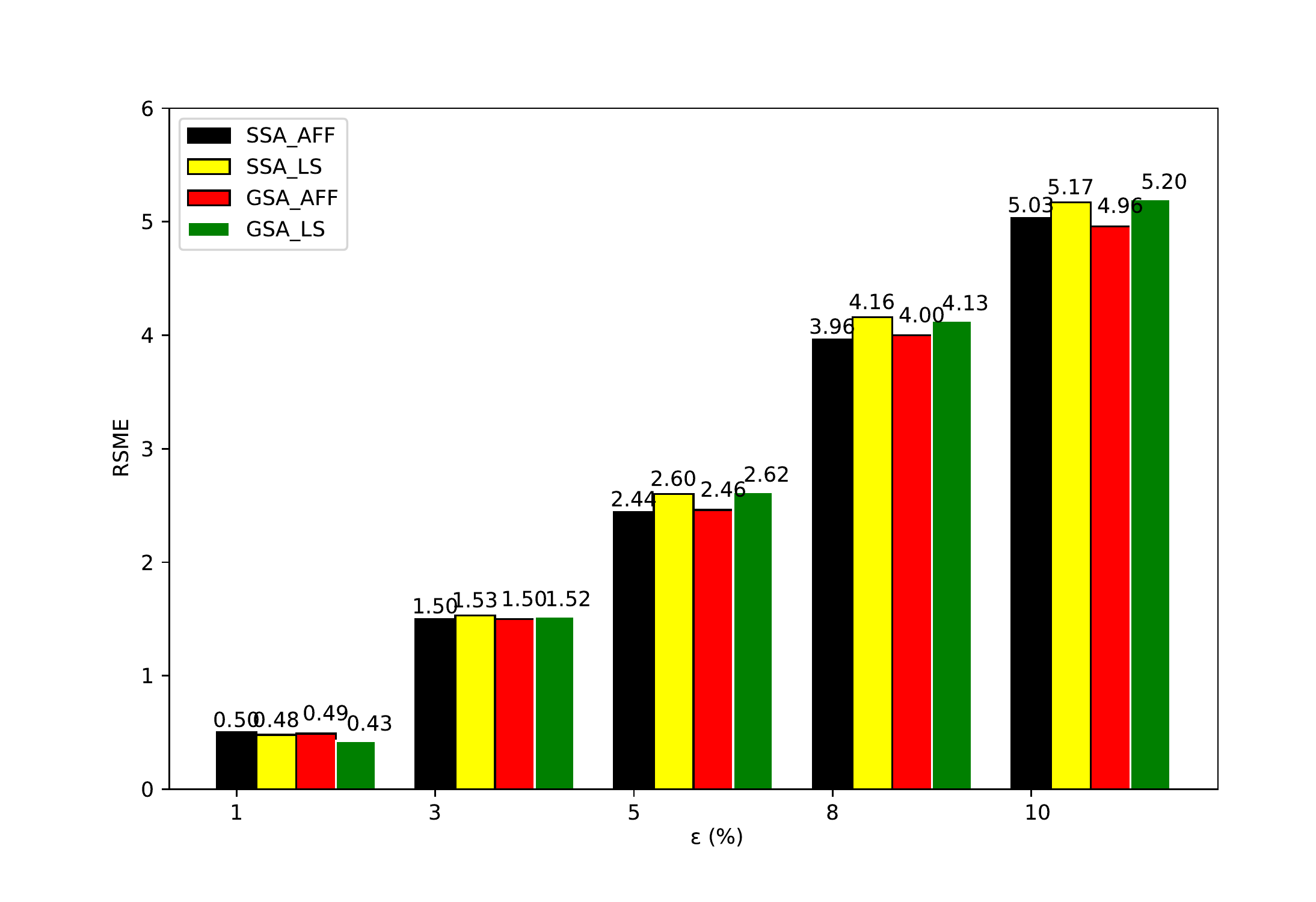}%
		\label{fig-4b}}
	\caption{ {The DSN\_ratio  and mean RMSE  results of the proposed dominant dataset selection methods with different $\varepsilon $ on DS0.}
	}
	\label{fig-4}
\end{figure}

\subsection{Effects of Parameter ${\delta }$ on the Dominant Dataset Selection}
\label{exp-C}
We further analyze the effects of parameter ${\delta }$ on the dominant dataset selection.
Fig. \ref{fig-5}  illustrates the experimental results of the proposed methods  on the DS0 dataset under the ${(\varepsilon ,\delta )}$-solver constraint where the error ${\varepsilon }$ is fixed as  5\% and  ${\delta }$ is respectively set as 1\%, 3\%, 5\%, 8\%, and 10\%. The related experimental results are also listed in Table \ref{tab5} and \ref{tab6}, respectively.
As Fig. \ref{fig-5a} illustrates,  the lower DSN\_ratio with the different values of ${\delta }$  is also achieved by the proposed GSA method compared to other methods.
For SSA, the  dominant objects are randomly selected
when  ${\delta }$ is greater than $\varepsilon$. From Fig. \ref{fig-5b}, we can see that
SSA makes  the mean RMSE greater than  5\% when ${\delta }$ is  greater than ${\varepsilon }$. This is because
the number of dominant objects allowed to exceed the error  ${\varepsilon }$ is large based on SSA. It means that
there are  negative effects on the  reconstruction  accuracy of target objects using the  random selection in SSA when
 ${\delta }$ is greater than ${\varepsilon }$.
For GSA, the object with the minimum mean of RMSE are selected as the dominant object when  ${\delta }$ is greater than $\varepsilon$. From Fig. \ref{fig-5b}, we can observe that it makes  the mean RMSE smaller based on GSA other than SSA.  GSA meets the constraint of ${\varepsilon}=5\%$  with the different values of ${\delta }$.
The performance of GSA is better than that of SSA
under the ${(\varepsilon ,\delta )}$-solver constraint.
In addition,
the  greedy selection algorithm based on the affine linear correlation measure, i.e., GSA\_AFF, is the optimal proposed method as shown in Fig. \ref{fig-5b}.

We also conducted the aforementioned experiments on the DS1-DS4 datasets, but, for brevity, we only present the experimental results on the DS0 dataset.
Although the correlation between the   TSD  and the number of dominatant datasets in the DS0-DS4  datasets are different,  the experimental results on the DS1-DS4 dataset are consistent with that on the DS0 dataset, which confirms that the proposed GSA\_AFF method is better than other three methods.

\begin{table}[htbp]
\begin{center}
	\caption{THE DSN\_RATIO WITH DIFFERENT $\delta$.}
\begin{tabular}{lllll}
\toprule
$\delta $&
SSA{\_}AFF&
SSA{\_}LS&
GSA{\_}AFF&
GSA{\_}LS \\
\midrule
1{\%}&
71.70\%&
85.66\%&
64.60\%&
81.40\% \\
3{\%}&
70.40\%&
83.72\%&
63.37\%&
79.75\% \\
5{\%}&
81.60\%&
81.60\%&
62.21\%&
78.10\% \\
8{\%}&
65.20\%&
79.36\%&
59.69\%&
75.39\% \\
10{\%}&
63.95\%&
76.94\%&
58.23\%&
73.84\% \\
\bottomrule
\end{tabular}
\label{tab5}
\end{center}
\end{table}

\begin{table}[htbp]
\begin{center}
	\caption{THE MEAN RSME WITH DIFFERENT $\delta$.}
\begin{tabular}{lllll}
\toprule
$\delta $&
SSA{\_}AFF&
SSA{\_}LS&
GSA{\_}AFF&
GSA{\_}LS \\
\midrule
1{\%}&
3.20&
3.90&
2.53&
2.77 \\
3{\%}&
4.50&
6.60&
2.63&
3.06 \\
5{\%}&
5.70&
8.56&
2.76&
3.23 \\
8{\%}&
7.50&
11.40&
2.97&
3.60 \\
10{\%}&
8.60&
13.59&
3.15&
3.89\\
\bottomrule
\end{tabular}
\label{tab6}
\end{center}
\end{table}

\begin{figure}[htbp]
	\centering
	\subfigure[]{\includegraphics[width=0.480\linewidth]{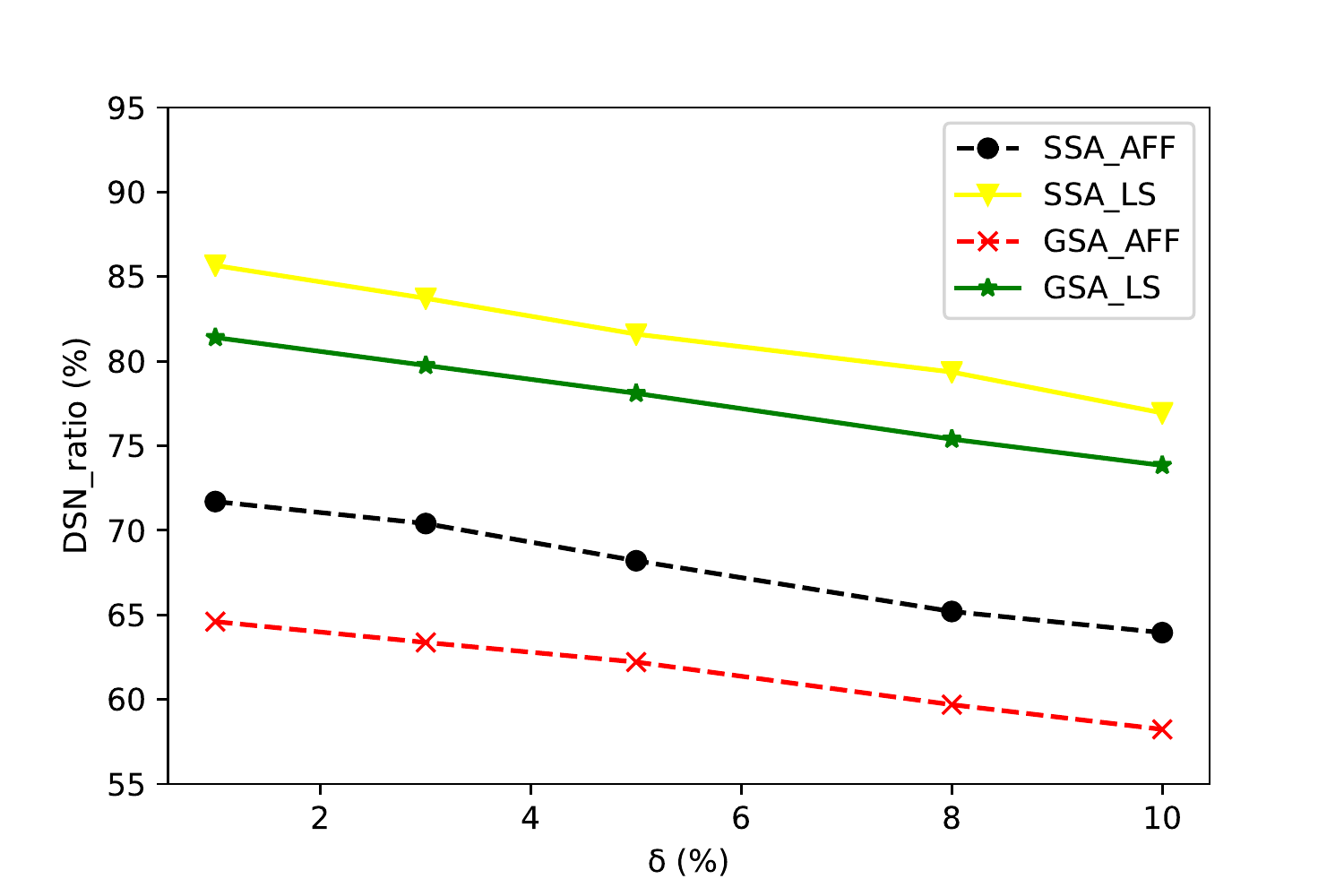}%
		\label{fig-5a}}
	\hfil
	\subfigure[]{\includegraphics[width=0.480\linewidth]{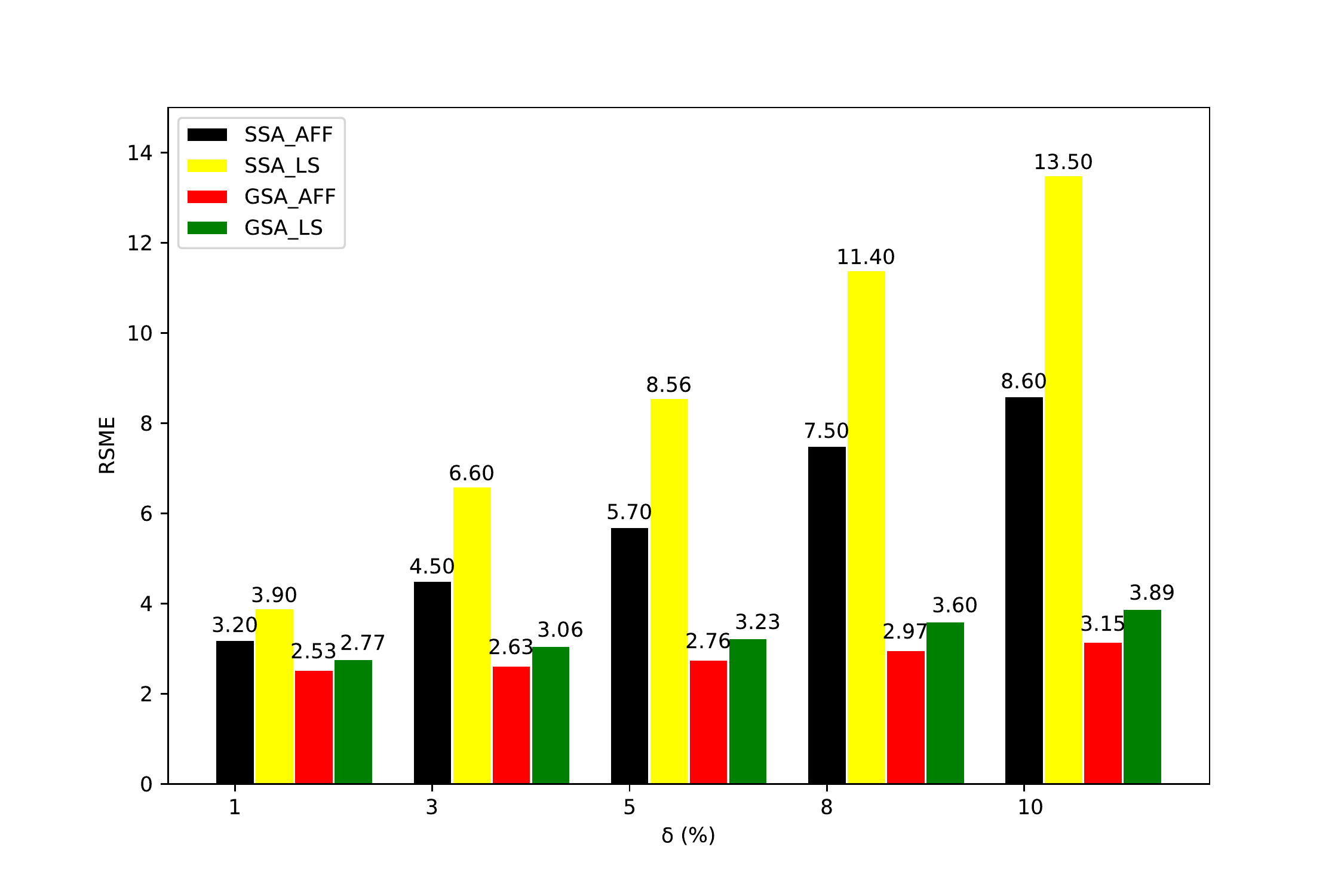}%
		\label{fig-5b}}
	\caption{ {The DSN\_ratio and mean RMSE results of the proposed dominant dataset selection methods with different ${\delta }$ on DS0}.
	}
	\label{fig-5}
\end{figure}

\subsection{Effects of Dataset Size on the Dominant Dataset Selection}

We further implement the proposed  dominant dataset selection methods on DS1, DS2, DS3, DS4  with ${\varepsilon }=5\% $. Table \ref{tab7} shows that it is consistent with the experimental results in Section \ref{exp-C}, that is, the proposed GSA\_AFF is the  optimal  method for the different size of datasets.
As illustrated in Fig. \ref{fig-6a}, the DSN\_ratios with the increasing size of datasets show a descend trend. It implies that the potential number of linear correlation relationships in TSD  objects rises with the increasing size of datasets. It relatively reduces  DSN\_ratios when Table \ref{tab7} is compared to Table \ref{tab3} with ${\varepsilon }=5\% $.
The number of target  objects corresponding to the central objects under the constraints increases to  improve the ability of representing target objects by the dominant objects. In Fig. \ref{fig-6b}, we can see that the mean RMSE of the different methods changes moderately in the different datasets.
From Table \ref{tab8}, we infer that  the different dataset size has little effect on the  reconstruction accuracy  based on the proposed methods.

\begin{table}[htbp]
\begin{center}
	\caption{THE DSN\_RATIO FOR DIFFERENT DATASETS BASED ON THE PROPOSED METHODS.}
\begin{tabular}{lllll}
\toprule
Dataset&
SSA{\_}AFF&
SSA{\_}LS&
GSA{\_}AFF&
GSA{\_}LS \\
\midrule
DS1&
29.67\%&
58.33\%&
24.00\%&
50.10\% \\
DS2&
22.33\%&
51.67\%&
19.00\%&
44.83\% \\
DS3&
19.56\%&
46.89\%&
16.22\%&
40.22\%\\
DS4&
16.75\%&
43.83\%&
14.17\%&
38.08\%\\
\bottomrule
\end{tabular}
\label{tab7}
\end{center}
\end{table}

\begin{table}[htbp]
\begin{center}
	\caption{THE RMSE FOR DIFFERENT DATASETS BASED ON THE PROPOSED METHODS.}
\begin{tabular}{lllll}
\toprule
Dataset&
SSA{\_}AFF&
SSA{\_}LS&
GSA{\_}AFF&
GSA{\_}LS \\
\midrule
DS1&
3.69&
5.10&
2.30&
2.69 \\
DS2&
3.71&
4.66&
2.35&
2.64 \\
DS3&
3.50&
5.35&
2.33&
2.64 \\
DS4&
3.52&
4.83&
2.30&
2.64\\
\bottomrule
\end{tabular}
\label{tab8}
\end{center}
\end{table}

\begin{figure}[htbp]
	\centering
	\subfigure[]{\includegraphics[width=0.480\linewidth]{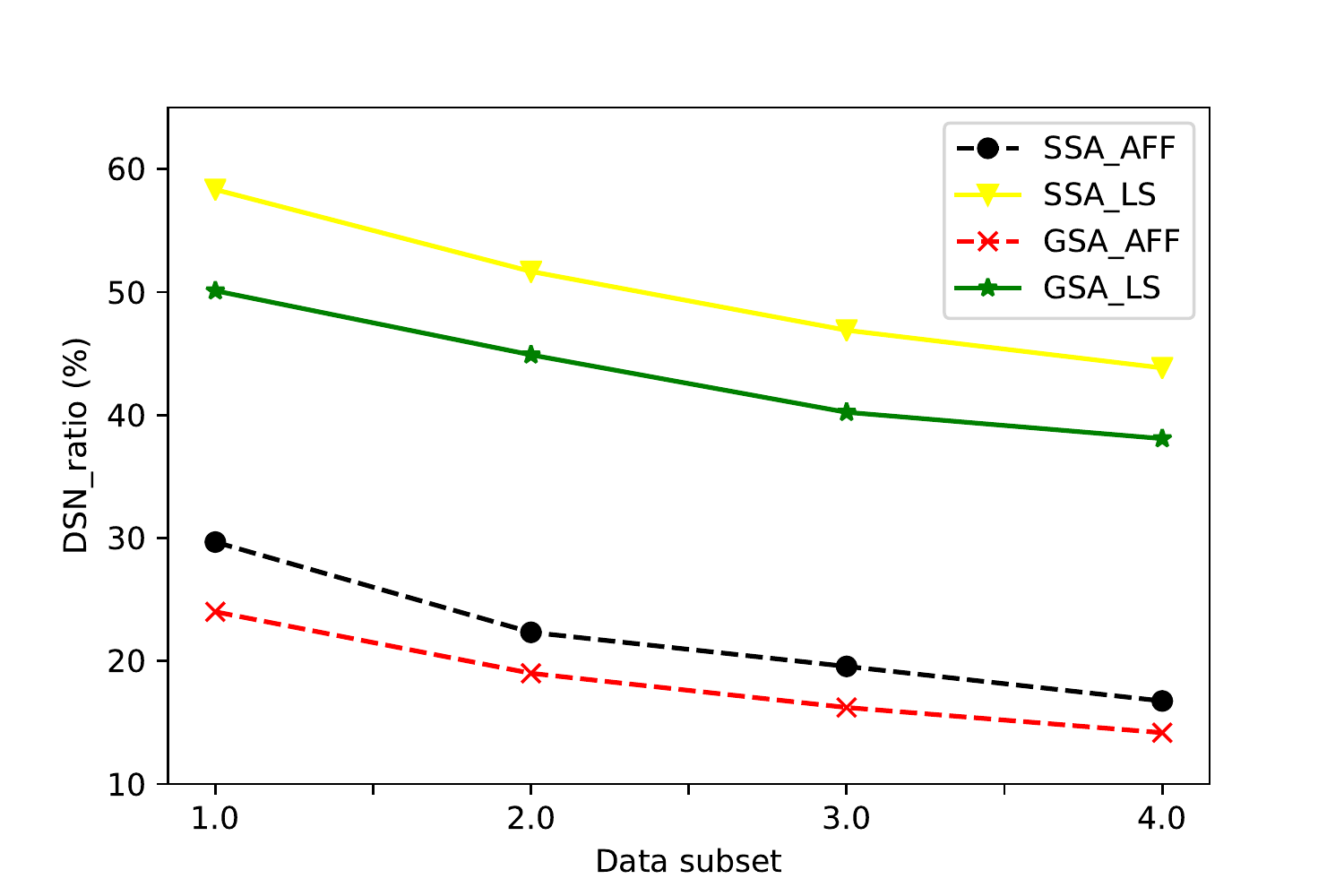}%
		\label{fig-6a}}
\hfil
	\subfigure[]{\includegraphics[width=0.480\linewidth]{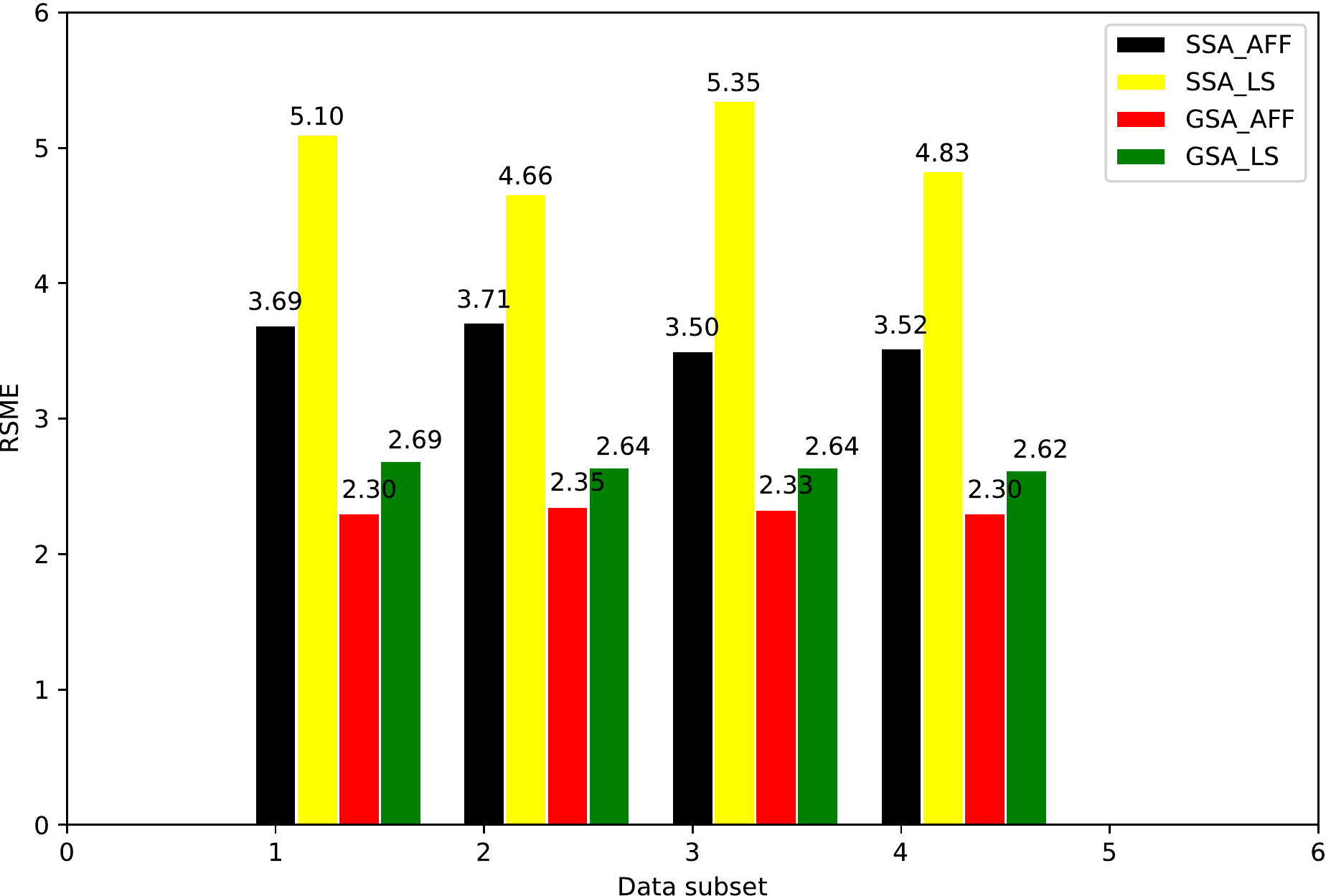}%
		\label{fig-6b}}
	\caption{  {The DSN ratio and mean RMSE results of the proposed dominant
dataset selection methods with the different dataset size.}
	}
	\label{fig-6}
\end{figure}

According to the proposed algorithm analysis in Section \ref{sec-6}, the time complexity of  SSA and GSA are {$O\left( {{n^3}} \right)$ and $O\left( {{n^4}} \right)$}, respectively. In the experiment, we further test the time consumption of the proposed methods. The simulations are implemented using the earlier computing
environment. Table \ref{tab-9} lists the average time consumptions of SSA\_AFF, SSA\_LS, GSA\_AFF, and GSA\_LS  with the constraints of ${\varepsilon =5\%,\delta=5\%}$ on the DS1-DS4 datasets.
For SSA, the time consumption based on the AFF measure is less than the LS measure. For GSA, however, the time consumption  based on  the LS measure is less than the AFF measure. It shows that
the time consumption of SSA is less than that of GSA in general.  With the increasing dataset size, the time consumption of SSA  has the obvious advantage over GSA.
Therefore, to make the trade-off between  the time consumption and the reconstruction accuracy, SSA\_AFF  may also be worth considering for the dominant dataset selection on large-scale datasets.
\begin{table}[htbp]
	\caption{ THE AVERAGE TIME CONSUMPTION OF THE PROPOSED  METHODS (SECONDS)}
	\begin{center} 
		\begin{tabular}{lllll}
\toprule
		Dataset& SSA\_AFF&	SSA\_LS&	GSA\_AFF&	GSA\_LS  \\
		\midrule
		DS1&8.2&14.02 & 92.55&78.8 \\
		DS2&21.1&42.4&408.67&352.47 \\
		DS3&36.3&83.67&909.12&862.21 \\
		DS4&52.21&138.72&1650.76&1653.76\\	
		\bottomrule
\end{tabular}
	\end{center}
	\label{tab-9}
\end{table}

\subsection{Discussion}

Based on the affine relation model, the central objects and the corresponding target objects can be identified by the proposed SSA and GSA algorithms.
Assuming that the central objects are ${p_1},{p_2}, \ldots ,{p_k}$, the  corresponding target object datasets  are ${{S}}_{p_1},{{S}}_{p_2}, \ldots ,{{S}}_{p_k}$ and  the sizes of ${{S}}_{p_1},{{S}}_{p_2}, \ldots ,{{S}}_{p_k}$ are  $n_{p_1},n_{p_2}, \ldots ,n_{p_k}$. The  number distribution  of $n_{p_1},n_{p_2}, \ldots ,n_{p_k}$ is analyzed as follows:

\begin{itemize}
\item[$\bullet$]
For SSA, the  distribution of $n_{p_1},n_{p_2}, \ldots ,n_{p_k}$ is converged  roughly in descending order. It is determined by the  mutuality of the linear correlation distance between the target objects (namely,  when  $i \ne j$, ${\mathfrak{D}_f}\left( {x_i,x_j} \right) = {\mathfrak{D}_f}\left( {x_j,x_i} \right)$).
The central objects ${p_1},{p_2}, \ldots ,{p_k}$  are  selected in order based on SSA.
With the scanning to the end, there are less probabilities of the identified target objects that meet the requirements of the constraint.
Therefore, $n_{p_1},n_{p_2}, \ldots ,n_{p_k}$ are generally arranged in descending order.
The above experimental results  are shown as Fig. \ref{fig-7}.

\item[$\bullet$]
For GSA, the distribution of $n_{p_1},n_{p_2}, \ldots ,n_{p_k}$ is converged completely in descending order.
The greedy selection strategy for the central objects is executed by selecting the central object that supports the maximum number of target objects in order. As shown in Fig. \ref{fig-8}, it illustrates the sizes of target object datasets are  completely in  descending order based on GSA.
In addition, it  also indicates a case that the target objects are incompletely selected by SSA.
It implies  that the greedy selection of  GSA  is better than the sequential selection of SSA.

\item[$\bullet$]
According to the distribution of $n_{p_1},n_{p_2}, \ldots ,n_{p_k}$, there are a certain number of central objects without supporting the target objects. It shows that there is no linear correlation between central objects and any target objects. Therefore, when these objects are used as the central objects, there are no corresponding target objects that could be selected.
It needs to be further explored whether there are other  nonlinear relationships between target objects to reduce the number of the central objects in our future work.

\end{itemize}

For the sake of fair assessment, the test dataset in Fig. \ref{fig-7} and Fig. \ref{fig-8} are constructed by extracting  the power consumption data from other power supply bureaus that do not belong to  Table \ref{tab-2}.
Additionally, this test dataset is constructed as the same scale as DS0.
Although the different linear correlation relationships between the constructed test dataset and DS0,
the representative experiments lead to the similar results as shown in Fig. \ref{fig-7} and Fig. \ref{fig-8}.
\begin{figure}[htbp]
	\centering
	\subfigure[]{\includegraphics[width=0.480\linewidth]{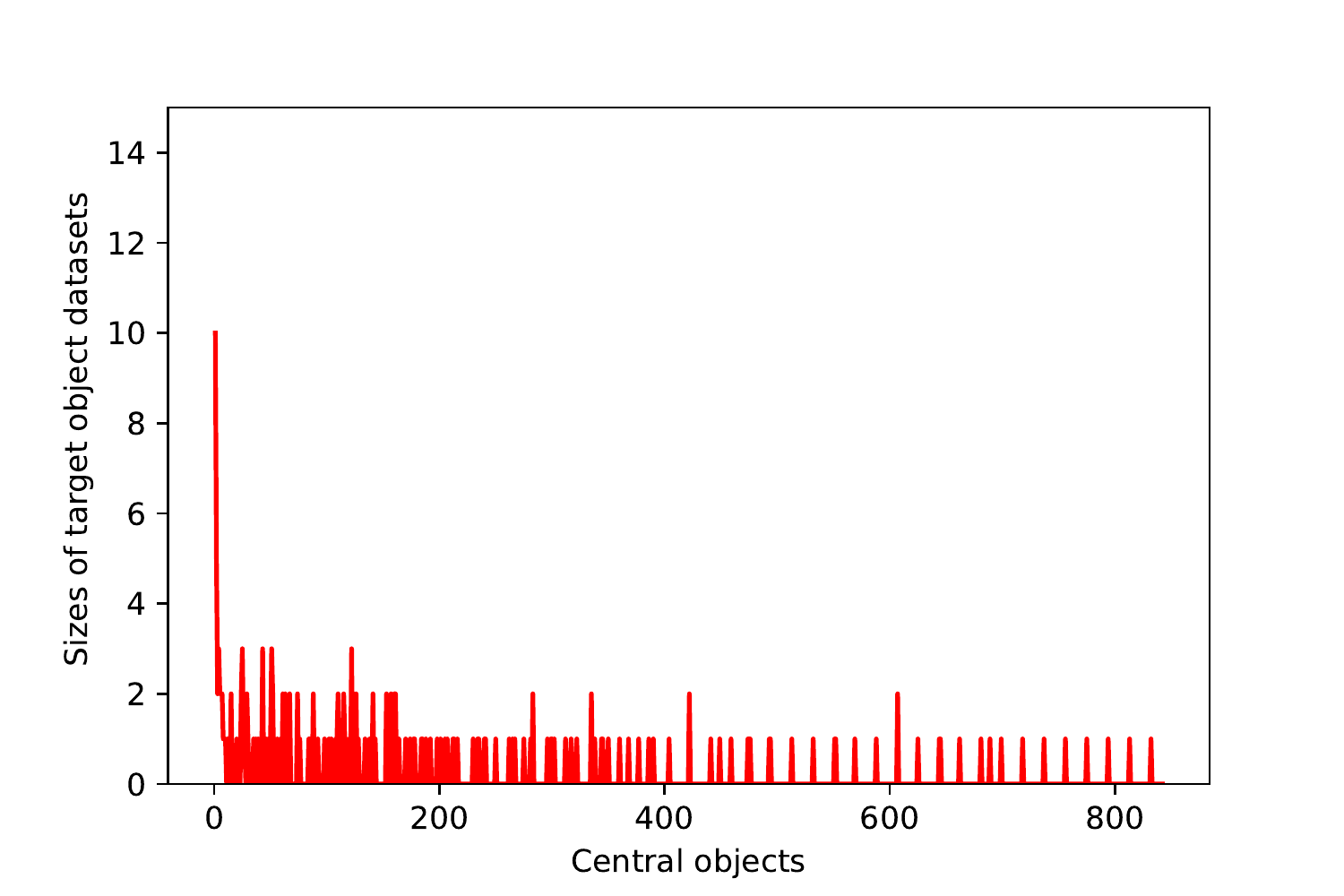}%
		\label{fig-7a}}
	\hfil
	\subfigure[]{\includegraphics[width=0.480\linewidth]{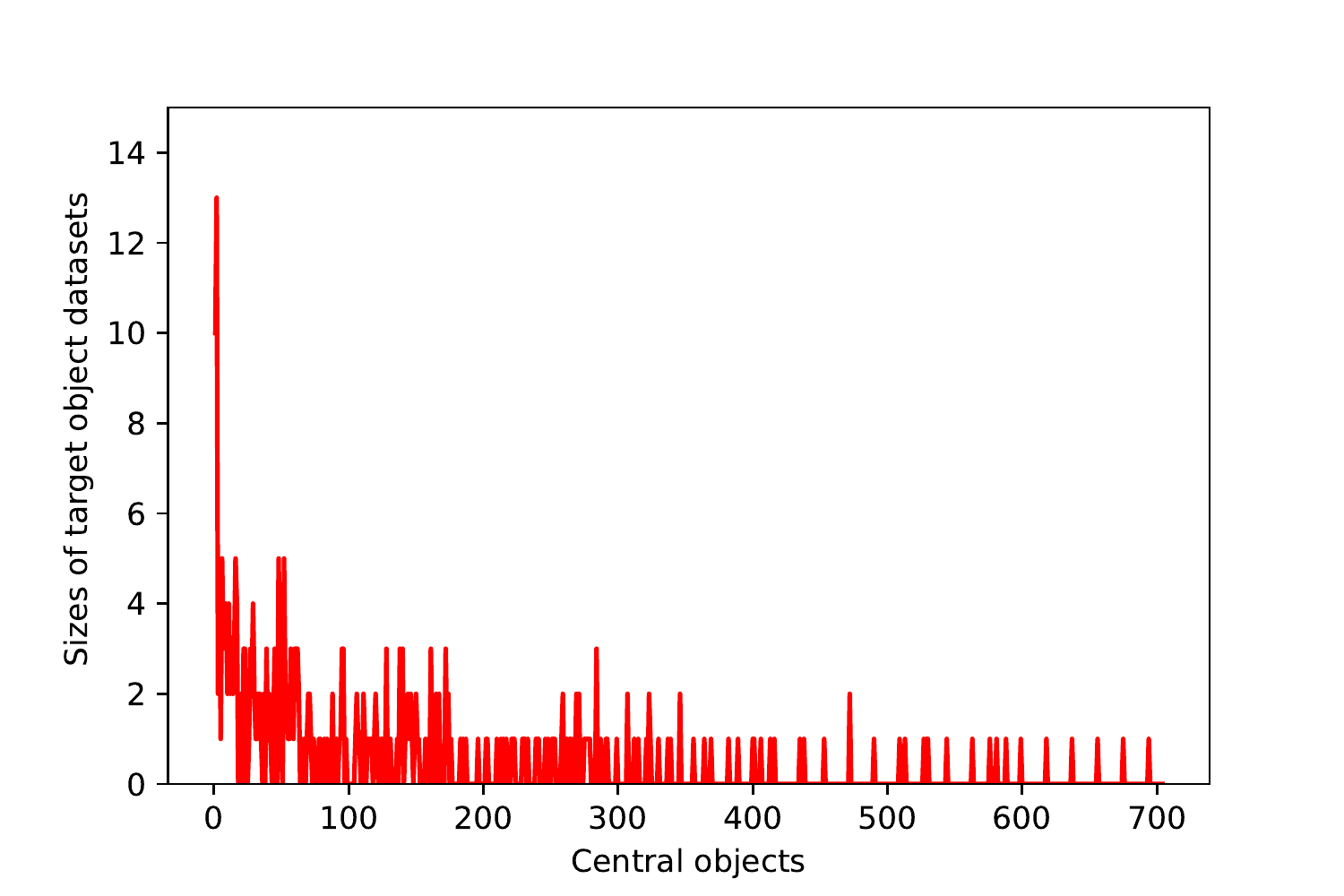}%
		\label{fig-7b}}
	\caption{ {The number distribution of  target objects based on (a) SSA\_LA (b) SSA\_AFF.}}

	\label{fig-7}
\end{figure}
\begin{figure}[htbp]
	\centering
	\subfigure[]{\includegraphics[width=0.48\linewidth]{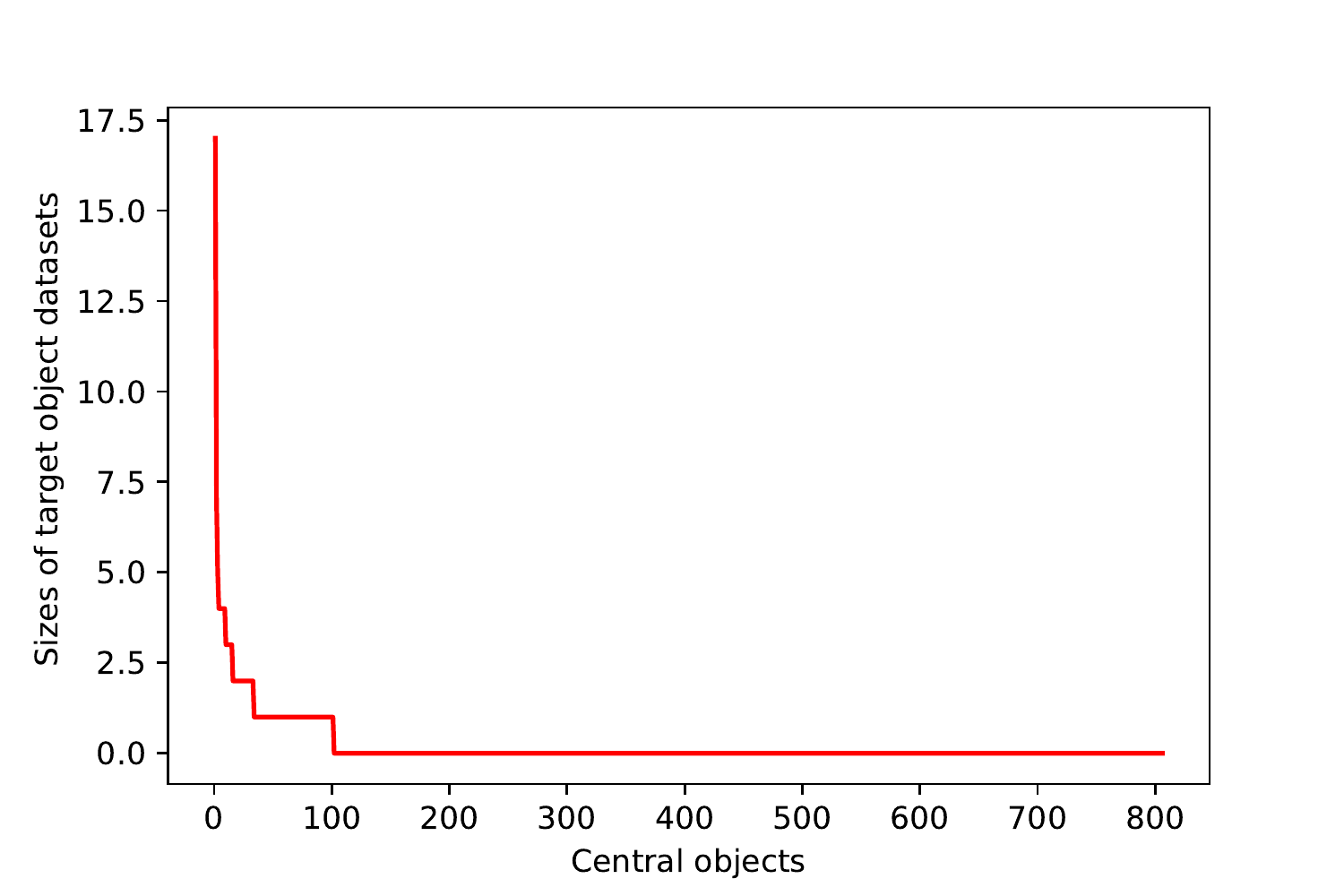}%
		\label{fig-8a}}
	\hfil
	\subfigure[]{\includegraphics[width=0.48\linewidth]{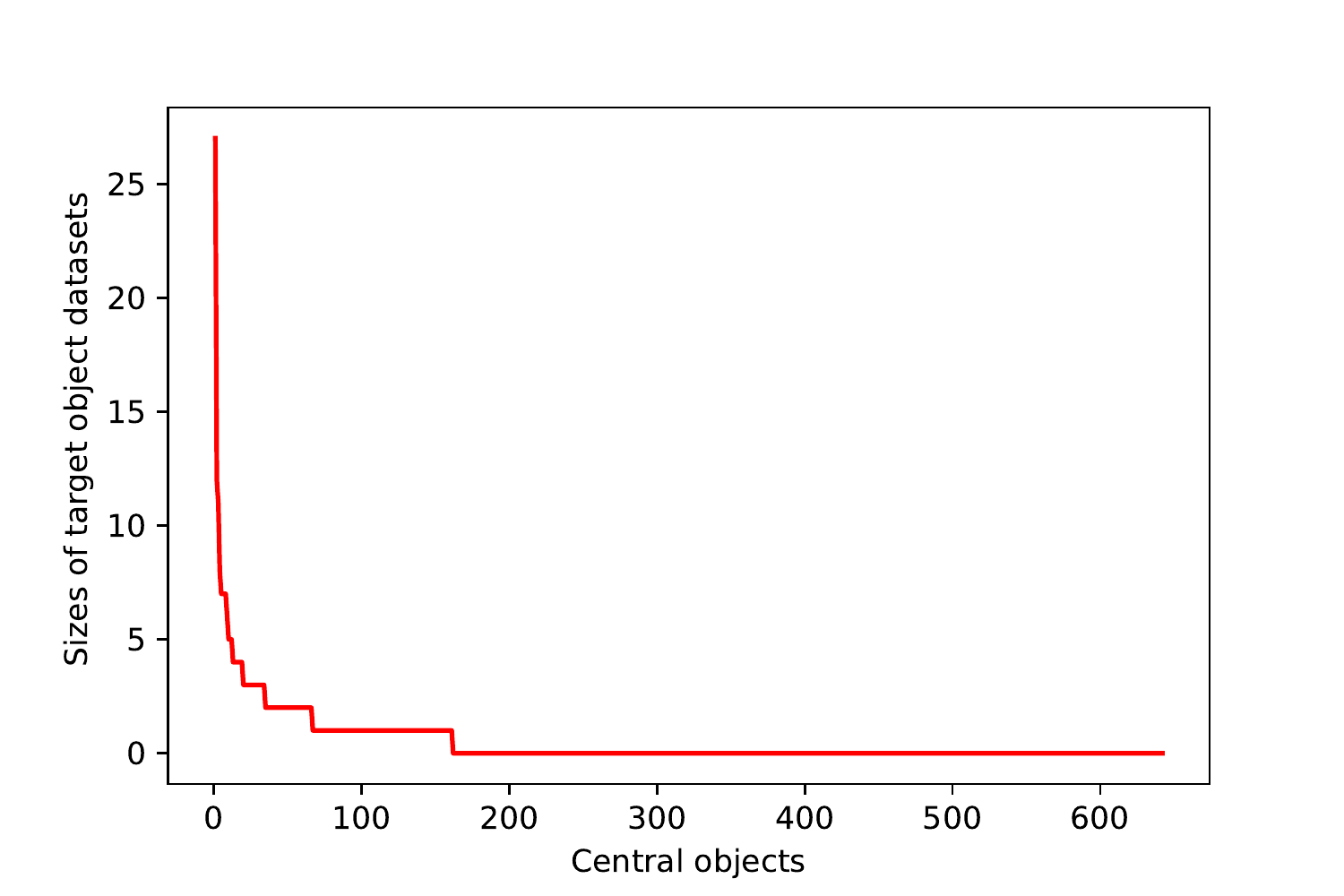}%
		\label{fig-8b}}
	\caption{ {The number distribution of target objects based on (a)
GSA\_LA (b) GSA\_AFF.}
	}
	\label{fig-8}
\end{figure}

\section{CONCLUSIONS}
\label{sec-8}

This paper studies how to extract the dominant dataset from electricity consumption time-series data.
We prove that the selection problem of the minimum dominant dataset is an NP-complete problem.
Based on the linear correlation relationship between time-series data objects, we present
a recursive affine transformation function to realize the efficient dominant dataset selection. In addition,
the  linear correlation distance is applied as the constraint condition for the dominant dataset selection. We further propose the dominant dataset selection algorithms based on the scanning strategy and the greedy strategy.
The analysis and experimental results show that the proposed algorithms have high performance in terms of effectiveness and efficiency.

In the future, we will reinvestigate the dominant dataset selection methods based on some linear and nonlinear relation models between time-series data, and further evaluate the proposed methods using the different types of time-series data derived from Internet of Things such as pressure, temperature, flow, etc.

\section{Acknowledgment}
This work is supported by Heilongjiang  Provincial Natural Science Foundation of China (Grant NO. F2016035) and
Science and Technology Project of State Grid Corporation of China (Grant NO. SGHL0000DKJS1900883).

\appendix

\begin{eqnarray*}
\resizebox{.95\hsize}{!}{$
\begin{aligned}
{R_{m + 1}} &= {\left( {P_{m + 1}^{'T} P_{m + 1}^'} \right)^{ - 1}} {P^'}_{m + 1}^T{S_{m + 1}}
\\
& = {\left( {{{\left( \begin{array}{l}
P{'_m}\\
p{'_{m + 1}}
\end{array} \right)}^T}\left( \begin{array}{l}
P{'_m}\\
p{'_{m + 1}}
\end{array} \right)} \right)^{ - 1}}{\left( \begin{array}{l}
P{'_m}\\
p{'_{m + 1}}
\end{array} \right)^T}\left( \begin{array}{l}
{S_m}\\
{s_{m + 1}}
\end{array} \right)
\\
&={\left(\left( {P_{m }^{'T}~ p_{m + 1}^{'T}} \right) \left( \begin{array}{l}
P{'_m}\\
p{'_{m + 1}}
\end{array} \right) \right)}^{-1}\left( {P_{m }^{'T} ~p_{m + 1}^{'T}} \right)\left( \begin{array}{l}
{S_m}\\
{s_{m + 1}}
\end{array} \right)
\\
&={\left( {P_{m }^{'T} P_{m }^'}+{p_{m+1}^{'T} p_{m+1}^'} \right)}^{-1}\left( {P_{m }^{'T} ~p_{m + 1}^{'T}} \right)\left( \begin{array}{l}
{S_m}\\
{s_{m + 1}}
\end{array} \right)
\\
&={\left( {\left( {P_{m }^{'T} P_{m }^'}\right)}\left(I_n+{\left( {P_{m }^{'T} P_{m }^'}\right)}^{-1}{p_{m+1}^{'T} p_{m+1}^'} \right)\right)}^{-1}\\
&~~~\times {\left( {P_{m }^{'T} S_{m }}+{p_{m+1}^{'T} s_{m+1}} \right)}
\\
&={\left( I_n+{\left( {P_{m }^{'T} P_{m }^'}\right)}^{-1}{p_{m+1}^{'T} p_{m+1}^'} \right)}^{-1}{\left( {P_{m }^{'T} P_{m }^'}\right)}^{-1}
\\&~~~\times {\left( {P_{m }^{'T} S_{m }}+{p_{m+1}^{'T} s_{m+1}} \right)}
\\
&={\left( I_n+{\left( {P_{m }^{'T} P_{m }^'}\right)}^{-1}{p_{m+1}^{'T} p_{m+1}^'} \right)}^{-1}
\\&~~~\times\left({\left( {P_{m }^{'T} P_{m }^'}\right)}^{-1}{{P_{m }^{'T} S_{m }}+{\left( {P_{m }^{'T} P_{m }^'}\right)}^{-1}{p_{m+1}^{'T} s_{m+1}} }\right)
\\
&= {\left( {{I_n} + {{\left( P_{m}^{'T}P'{_m} \right)}^{ - 1}}p_{m + 1}^{'T}p'{_{m + 1}}} \right)^{ - 1}}
 \\&~~~\times\left( {{R_m} + {{\left( {P'_m}^TP{'_m} \right)}^{ - 1}}{{p'}^T_{m + 1}}{s_{m + 1}}} \right)
\end{aligned}$}
\end{eqnarray*}

\ifCLASSOPTIONcaptionsoff
\newpage
\fi
\bibliographystyle{IEEEtran}
\bibliography{reference}

\end{document}